\documentclass[sigconf]{acmart}   

\acmConference[To appear in ASIA CCS]{ACM Asia Conference on Computer and Communications Security}
\setcopyright{none} 
\acmYear{2022}

\usepackage{soul}
\usepackage{url}
\usepackage[switch]{lineno}
\usepackage[utf8]{inputenc}
\usepackage{graphicx}
\usepackage{amsmath}
\usepackage{amsthm}
\usepackage{amssymb}
\usepackage{algorithmicx}
\usepackage{algpseudocode}
\usepackage{enumitem}
\usepackage{moresize}
\allowdisplaybreaks

\usepackage{epsfig}
\usepackage{amsfonts}
\usepackage{amssymb}
\usepackage{multirow}
\usepackage{multicol}
\usepackage{subfig}
\usepackage{comment}
\usepackage{color}
\usepackage[utf8]{inputenc}

\usepackage{balance}

\usepackage{filecontents}

\usepackage{thm-restate,thmtools}

\makeatletter
\algrenewcommand\ALG@beginalgorithmic{\small}
\makeatother

\makeatletter  
\newif\if@restonecol  
\makeatother

\usepackage[linesnumbered,ruled,vlined]{algorithm2e}
\usepackage{algpseudocode}  
\usepackage{amsmath}  
\renewcommand{\algorithmicrequire}

\newcommand{\myparatight}[1]{\noindent{\bf {#1}:}~}

\renewcommand{\algorithmicrequire}{ \textbf{Input:}} 

\AtBeginDocument{%
  \providecommand\BibTeX{{%
    \normalfont B\kern-0.5em{\scshape i\kern-0.25em b}\kern-0.8em\TeX}}}

\setcopyright{rightsretained}

\usepackage{etoolbox}
\makeatletter
\patchcmd{\maketitle}{\@copyrightpermission}{
\begin{minipage}{0.4\columnwidth}
\href{http://creativecommons.org/licenses/by/4.0/}{\includegraphics[width=0.5\textwidth]{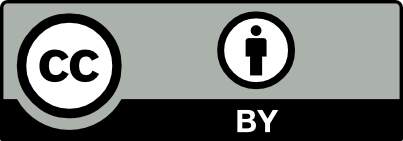}}
\end{minipage}\hfill
\begin{minipage}{0.6\columnwidth}
\end{minipage}\hfill

\href{http://creativecommons.org/licenses/by/4.0/}{This work is licensed under a Creative Commons Attribution International 4.0 License.}}

\begin{document}

\vspace{3pt}
\copyrightyear{2022}
\acmYear{2022}
\acmConference[ASIA CCS '22]{Proceedings of the 2022 ACM Asia Conference on Computer and Communications Security}{May 30-June 3, 2022}{Nagasaki, Japan}
\acmBooktitle{Proceedings of the 2022 ACM Asia Conference on Computer and Communications Security (ASIA CCS '22), May 30-June 3, 2022, Nagasaki, Japan}
\acmDOI{10.1145/3488932.3517398}
\acmISBN{978-1-4503-9140-5/22/05}

\makeatother


\title{GraphTrack: A Graph-based Cross-Device Tracking Framework}

\author{Binghui Wang$^{1}$, Tianchen Zhou$^{2}$, Song Li$^{3}$, Yinzhi Cao$^{3}$, Neil Gong$^{4}$} 
\affiliation{
$^{1}$Illinois Institute of Technology  
$^{2}$The Ohio State University
$^{3}$Johns Hopkins University
$^{4}$Duke University
\country{USA}\\
bwang70@iit.edu, 
zhou.2220@osu.edu,
\{lsong18, yinzhi.cao\}@jhu.edu,
neil.gong@duke.edu
}


\begin{abstract}
Cross-device tracking has drawn growing attention from both commercial companies and the general public because of its privacy implications and applications for user profiling, personalized services, etc.
One particular, wide-used type of cross-device tracking is to leverage browsing histories of user devices, e.g., characterized by a list of IP addresses used by the devices and  domains visited by the devices. 
However,  existing browsing history based methods have three drawbacks.  First, they cannot capture latent correlations among IPs and domains. Second, their performance degrades significantly when labeled device pairs are unavailable. Lastly, they are not robust 
to 
uncertainties in linking browsing histories to devices. 

We propose \emph{GraphTrack}, 
{a graph-based cross-device tracking framework,} 
to track users across different devices by correlating their browsing histories. 
Specifically,  we propose to model the complex interplays among IPs, domains, and devices as graphs and capture the latent correlations between IPs and between domains. We construct graphs that are robust to uncertainties in linking browsing histories to devices. Moreover, we adapt \emph{random walk with restart} to compute similarity scores between devices based on the graphs. 
GraphTrack leverages the similarity scores to perform  cross-device tracking. GraphTrack does not require labeled device pairs and can incorporate them if available.
We evaluate GraphTrack on two real-world datasets, i.e., a publicly available mobile-desktop tracking dataset (around 100 users) and a multiple-device tracking dataset (154K users) we collected.
    Our results show that GraphTrack substantially outperforms the state-of-the-art on both datasets.
   
\end{abstract}



\maketitle

\newcommand{\reducedrefsize}{\fontsize{8}{8}\selectfont}
{\reducedrefsize \textbf{ACM Reference Format:}\\
{Binghui Wang, Tianchen Zhou, Song Li, Yinzhi Cao, Neil Gong. 2022. GraphTrack: A Graph-based Cross-Device Tracking Framework. In \textit{Proceedings of the 2022 ACM Asia Conference on Computer and Communications Security (ASIA CCS '22), May 30-June 3, 2022, Nagasaki, Japan}. 
ACM, New York, NY, USA, 15 pages. \url{https://doi.org/10.1145/3488932.3517398}}}

\section{Introduction}
Cross-device tracking---a technique used to identify 
whether various devices, such as mobile phones and desktops, have common owners---has drawn much attention of both commercial companies and the general public. 
For example, Drawbridge~\cite{drawbridge}, an advertising company, 
goes beyond traditional device tracking to identify devices belonging to the same user.
 Due to the {increasing demand for cross-device tracking} and corresponding privacy concerns,  
 the U.S. Federal Trade Commission hosted a workshop~\cite{ftccdt} in 2015 and released a staff report~\cite{ftcreport} about cross-device tracking and industry regulations in early 2017. 
 The growing interest in cross-device tracking is highlighted by 
 the privacy implications associated with tracking and the applications of tracking for user profiling, personalized services, and user authentication. 
 For example, a bank application can adopt cross-device tracking as a part of multi-factor authentication to increase account security. 

Generally speaking, cross-device tracking mainly leverages \emph{cross-device IDs}, \emph{background environment}, or \emph{browsing history} of the devices. For instance,  cross-device IDs may include a user's email address or username, which are not applicable when users do not register accounts or do not login. Background environment (e.g., ultrasound~\cite{mavroudis17:ultrasound}) also cannot be applied when devices are used in different environments such as home and workplace. {Therefore, we focus on browsing history based cross-device tracking~\cite{zimmeck2017privacy,deviceGraphKDD17}}. 

Specifically, browsing history based tracking utilizes
source and destination pairs---e.g., the client IP 
address 
and
the destination website's domain---of users' browsing
records to correlate different devices of the same user. 
 Several browsing history based cross-device tracking methods~\cite{cao2015recovering,zimmeck2017privacy,deviceGraphKDD17}  have been proposed. For instance, IPFootprint~\cite{cao2015recovering} uses supervised learning to analyze the IPs commonly used by devices.   Zimmeck et al.~\cite{zimmeck2017privacy} proposed a {supervised}  method that achieves state-of-the-art performance. In particular, their method computes a  similarity score via Bhattacharyya coefficient~\cite{wang2013uniquely} for a pair of devices based on the {common} IPs  and/or domains visited by both devices. Then, they use the similarity scores to track devices. We call the method \emph{BAT-SU} since it uses the Bhattacharyya coefficient, where the suffix ``-SU'' indicates that the method is supervised. DeviceGraph~\cite{deviceGraphKDD17} is an unsupervised method that models devices as a graph based on their IP colocations (an edge is created between two devices if they used the same IP) and applies community detection for tracking, i.e., the devices in a community of the graph belong to a user. 
 
However, existing browsing history based  methods have three
major limitations. 
First, they cannot capture the latent correlations among domains and IPs in the browsing history. Let us first look at the domain.  Suppose both Facebook and Twitter are frequently visited by a large amount of devices, and thus there exists a certain latent correlation between them (in this example, both of them are social media sites).  Assume a user visits Facebook frequently on one device but visits Twitter frequently on another device. The two devices could have a large similarity score because they both frequently visit sites that have latent correlations. 
However, state-of-the-art method like BAT-SU~\cite{zimmeck2017privacy} would compute a very low similarity score for the two devices because BAT-SU only leverages the common domains visited by both devices to compute similarity scores.  
Likewise, IP addresses of devices could have latent correlations. 
For example, say many users travel between a dormitory and a campus in a daily basis. Thus, a dormitory IP on one device and a campus IP on another should produce a large similarity score between the two devices, but existing methods fail to do so. 
Second, existing methods have limited performance when applied to the scenarios where labeled device pairs are unavailable, such as a third-party tracker scenario. 
  A tracker is a \emph{third-party tracker} if the tracker tracks users who visit other parties' web services. For instance, the ad network \texttt{Atlas} is a third-party tracker which can track users who visit the web services 
  using \texttt{Atlas}. It is often challenging for a third-party tracker to obtain labeled device pairs because users are usually unaware of and do not interact with third-party trackers. 
   Third, existing methods are not robust to uncertainties that exist in device identification, or called single-device tracking. For example, when  single-device tracking adopts browser fingerprinting, uncertainties or errors could happen when  linking browsing histories to devices~\cite{fingerprinting1,fingerprinting2,Acar:2014:WNF:2660267.2660347,Acar:2013:FDW:2508859.2516674,laperdrix2016beauty,Boda:2011:UTW:2341491.2341497,fmeasure}.  Existing cross-device tracking methods are not robust to such uncertainties.

We propose \emph{GraphTrack}, a graph-based framework, to perform cross-device tracking. 
GraphTrack overcomes  above limitations of existing cross-device tracking methods. 
First, we leverage the complex interplays between devices and browsing histories of many users to capture the latent correlations. Our intuition is that if many devices visit two domains (or use two IPs), then there could be a latent relationship between them. To capture such intuition, we propose to model  the complex interactions between IPs, domains, and devices as graphs; and we adapt \emph{random walk with restart} (RWwR) on the graphs to compute similarity scores between devices, which captures the latent correlations between domains/IPs. Specifically, we propose an \emph{IP-Device graph} to model the interactions between IPs and devices, and a \emph{Domain-Device graph} to model interactions between domains and devices. We distinguish IPs and domains because they are different data types. 
 In the IP-Device (or Domain-Device) graph, a node means  an  IP (or domain) and an edge means the corresponding device used the corresponding IP (or visited the corresponding domain).  We note that BAT-SU essentially uses the number of common neighbors (with certain normalization) of two devices in our proposed graphs  to compute the similarity score between the two devices. GraphTrack leverages RWwR, which can better capture the graph structure, to compute similarity scores.   
 
Second, to be robust to uncertainty in single-device tracking, we further model the weight of an edge as the number of times that the device used the IP (or visited the domain). For instance, suppose a device visited a domain multiple times; once a single-device tracker links a majority of them to the device, the associated edge still has a large weight. However, if a device did not visit a certain domain, but the single-device tracker occasionally links the domain with the device, then the weight of the corresponding edge is small. Thus, the incorrectly linked domain has a small impact on the overall structure of the Domain-Device graph.

Third, GraphTrack leverages the similarity scores computed based on the graphs to perform cross-device tracking. 
Suppose a user has $K$ devices. Then, any pair of the $K$ devices is likely to have a large similarity score. Based on this intuition, GraphTrack constructs a device \emph{similarity graph}, where a node is a device and two nodes are connected if their similarity score is large enough. Then, GraphTrack identifies the devices in a \emph{clique} in the similarity graph belong to the same user.   
When no labeled device pairs are available, GraphTrack adds an edge to two devices in the similarity graph if and only if one device is among the top-$(K-1)$ most similar devices of the other device and vice versa. When  labeled device pairs are available,  GraphTrack uses them to learn a threshold of similarity score and adds an edge to two devices in the similarity graph if their similarity score is no less than the threshold.

We compare GraphTrack with state-of-the-art methods
 using two real-world datasets: one for mobile-desktop tracking~\cite{zimmeck2017privacy} ($\sim$ 100 users), where all users have at most 
one mobile/desktop device; 
and the other for multiple-device tracking (154K users), where users have 2-5 devices (we don't know the device types). 
{First, when labeled device pairs are available, GraphTrack consistently outperforms compared supervised methods. 
For instance, when 20\% users are labeled,
 GraphTrack outperforms 
BAT-SU~\cite{zimmeck2017privacy} and IPFootprint~\cite{cao2015recovering} for mobile-desktop tracking by 0.15 and 0.25 in Accuracy; and for multiple-device tracking by 0.1 and 0.13 in Accuracy, respectively. 
Second, when no labeled device pairs are available, GraphTrack consistently outperforms compared unsupervised methods. 
For instance, GraphTrack outperforms the unsupervised version of BAT-SU by 0.1 and 0.08 and DeviceGraph~\cite{deviceGraphKDD17} by 0.64 and 0.22 in Accuracy on the two datasets, respectively.}
Third, GraphTrack is robust to uncertainty in single-device tracking. 
For instance, 
GraphTrack's Accuracy is still 0.72 when the single-device tracker incorrectly links 10\% of each device's browsing history to random devices. 
Fourth, GraphTrack takes less than 40ms to track each device in our multiple-device dataset, demonstrating its practicability as a real-world tracker.
Our results show that cross-device tracking poses privacy threats to more users than previously thought. 
Our key contributions are summarized as follows:
\begin{itemize}[leftmargin=*]
\item We propose \emph{GraphTrack}, a graph-based framework, for cross-device tracking. 
  GraphTrack is applicable with or without labeled device pairs, robust to uncertainty in single-device tracking, and practical as a real-world tracker.
  
\item We propose to model complex interplays between IPs, domains, and devices as graphs. Moreover, we adapt RWwR to analyze the structure of the graphs and capture latent correlations among IPs and domains. 

\item We evaluate GraphTrack and compare it with state-of-the-art methods on two real-world datasets. Our results  show that GraphTrack 
consistently outperforms these methods.

\end{itemize}

\vspace{-2mm}

\section{Related Work}
\label{relatedwork}

\noindent {\bf Single-device tracking:} 
It refers to techniques that are used to identify a single device, such as a desktop, a mobile phone, or a tablet.   Prior work on single-device tracking can be roughly classified into two categories: \emph{cookie or super-cookie based} and \emph{browser fingerprinting}.   
  First, Roesner et al.~\cite{nsdi} surveyed and measured top Alexa websites and identified a significant number of trackers in the wild. Lerner et al.~\cite{lerner2016} conducted an archaeological study of web tracking using Internet time machine to understand the evolution of tracking from 1996 to 2016. Metwalley et al.~\cite{metwalley2015unsupervised} measured web tracking using an unsupervised method.  Other than measuring the significance of tracking, some research work~\cite{krishnamurthy2006generating,krishnamurthy2009privacy,krishnamurthy2011privacy,krishnamurthy2008characterizing,sok,sanchez2015tracking} studied the privacy implication of web tracking, such as business model and the leak of email addresses and user names. 
  Second, several works~\cite{fingerprinting1,fingerprinting2,Acar:2014:WNF:2660267.2660347,Acar:2013:FDW:2508859.2516674,laperdrix2016beauty,Boda:2011:UTW:2341491.2341497,fmeasure} performed measurement studies on browser fingerprinting, a second-generation web tracking technique that utilizes browser features such as number of plugins, fonts, and user agents. 
  Fifield et al.~\cite{fifield2015fingerprinting} focused on a specific feature of browser, i.e., fonts, and proposed to use a subset of fonts for browser fingerprinting.  Mowery et al.~\cite{mowery2012pixel} proposed to use canvas, a HTML5 feature, for fingerprinting.  Mulazzani et al.~\cite{mulazzani2013fast} and Mowery et al.~\cite{mowery2011fingerprinting} fingerprinted browsers using features of JavaScript engine.  Nakibly et al.~\cite{nakibly2015hardware} proposed 
  several 
  tracking techniques using features from hardware including microphone, motion sensor, and GPU.  
  Cao et al.~\cite{ndss17cao} extended existing browser fingerprinting techniques to be cross-browser, e.g., among IE, Firefox, and Chrome.

Single-device tracking is the basis of cross-device tracking, 
which identifies individual devices first and then links devices together.  
Cross-device tracking also goes beyond single-device tracking, because it can identify a common user behind different devices.

\noindent {\bf Cross-device tracking:} 
It is a relatively new research area and refers to techniques 
used to identify devices belonging to the same user. 
 In 2015, Drawbridge released a challenge~\cite{drawbridgechallenge} to the research community about IP based cross-device tracking. Then,  multiple research papers~\cite{walthers2015learning,selsaas2015affm,anand2015machine,diaz2015cross,landry2015multi,kim2015connecting,kejela2015cross,cao2015recovering} on IP based cross-device tracking were published. 
{Among these methods, IPFootprint~\cite{cao2015recovering} achieves state-of-the-art performance. IPFootprint leverages a learning-to-rank method called RankNet~\cite{burges2005learning} to analyze the set of IPs together with the frequencies a device used these IPs.}
 Zimmeck et al.~\cite{zimmeck2017privacy} proposed a supervised method called BAT-SU, which {leverages both IPs and domains} and achieves state-of-the-art performance. 
 These supervised methods require the tracker to manually label a large number of device pairs.
 {Malloy et al.~\cite{deviceGraphKDD17} proposed an unsupervised method (we call DeviceGraph) based on community detection. Specifically, DeviceGraph first constructs a {device graph} based on IP colocations, i.e., a node is a device and an edge is created between two devices if they used the same IP. Then, DeviceGraph detects communities in the device graph and predicts that devices in a community belong to the same user.
  These methods suffer from three major limitations as 
  we discussed in Introduction.} 
 {Note that several companies~\cite{drawbridge,criteo} were also reported to use graph analysis for cross-device tracking. However, they did not disclose their technical details.}
Solomos et al.~\cite{solomos2019talon} proposed to audit the cross-device tracking ecosystem in an automated way. This work aims to understand the inner workings of the cross-device tracking mechanics and is orthogonal to our work. 

Apart from browsing history based cross-device tracking, ultrasound was also used 
to link different devices~\cite{mavroudis16:ultrasound,mavroudis17:ultrasound} when they are positioned in the same environment. By contrast, browsing history based cross-device tracking can link two devices even when they are in different environments, such as at home and in the workplace. Brookman et al.~\cite{brookman2017cross} conducted a measurement  study about cross-device tracking in the wild. They found that websites often share a large amount of data with third parties that could allow them to track users’ 
devices.

\noindent{\bf Defense against single-device tracking:}  As tracking violates user privacy, many methods have been proposed to defend against device tracking. Existing defenses focus on single-device tracking and can be classified into two categories: defense against cookie and super-cookie based tracking and defense against browser fingerprinting. First,  
ShareMeNot~\cite{nsdi}, private browsing mode~\cite{private_browsing,Xu:2015:UPB:2810103.2813716}, TrackingFree~\cite{trackingfree}, and Meng et al.~\cite{Meng:2016:TEF:2872427.2883034} are examples of defending against cookie and super-cookie based tracking.  The key idea of such approaches is to isolate tracking entities into different  units and prevent tracking.  
Second, 
FuzzyFox~\cite{fuzzyfox} and PriVaricator~\cite{Nikiforakis:2015:PDF:2736277.2741090} proposed adding noise to browsers so that the fingerprint will change all the time.  On the contrary, DeterFox~\cite{deterfox}  and Tor Browser~\cite{perry2011design} 
 normalize the outputs so that the fingerprinting results remain the same across different devices.  Although many defenses have been proposed, device and cross-device tracking is still popular in the current Internet landscape.

\vspace{-1mm}
 \section{Problem Definition}

\myparatight{Threat model}
We consider both \emph{first-party tracker} and \emph{third-party tracker}. 
A first-party tracker tracks the users who visit its web services. For instance, Facebook is a first-party tracker when tracking users who use Facebook. 
A third-party tracker tracks users who visit other parties' web services. Example third-party trackers could be an ad network like \texttt{Atlas}, a social network add-on like \texttt{Facebook like button}, and a third-party JavaScript library like Google Analytics. Such third-party trackers can collect users' browsing history via, e.g., the referer header upon the websites that embed such trackers. According to existing surveys~\cite{nsdi}, 
more than 
$90\%$ of Alexa Top websites include at least one third-party tracker, and some third-party trackers even collaborate with each other. Note that some third-party tracker like \texttt{Facebook like button} also serves as the first party. 

Suppose a tracker 
has collected a large amount of  $(IP, Domain)$ pairs from many devices at different time, where a pair $(IP, Domain)$ from a device means that the device once visited $Domain$ using the $IP$ address.  
 The tracker first links the pairs $(IP, Domain)$ to devices, which is known as \emph{single-device tracking}. Various techniques---such as cookies or super-cookies~\cite{nsdi}, browser fingerprinting~\cite{fingerprinting1,fingerprinting2,Acar:2014:WNF:2660267.2660347,Acar:2013:FDW:2508859.2516674,laperdrix2016beauty,Boda:2011:UTW:2341491.2341497,fmeasure}, and cross-browser fingerprinting~\cite{Boda:2011:UTW:2341491.2341497,ndss17cao}---can be used to perform single-device tracking. 
 In this paper, we assume that single-device tracking has been done by existing works. 
 
After single-device tracking, we have a set of devices and a \emph{browsing history} $\{(IP_1, Domain_1), \cdots, (IP_k, Domain_k)\}$ for each device. Note that the same pair $(IP, Domain)$ could appear multiple times in the browsing history of a device.
The tracker's goal is to divide the devices into disjoint groups, where each group of devices belong to the same user (a group could have just one device). 
After cross-device tracking, the tracker can combine the IPs accessed and domains visited by the devices to better profile a user, e.g., infer the user's demographics and interests with higher accuracies~\cite{zimmeck2017privacy}, which in turn helps deliver targeted advertisements to the user. Formally, we define the \emph{cross-device tracking problem} as follows:
\vspace{-1mm}
 \begin{definition}[Cross-device tracking]
 Suppose we are given a set of $n$ devices $\mathcal{D}=\{D_1, D_2, \cdots, D_n\}$, and a browsing history of each device produced by single-device tracking. 
 Cross-device tracking is to divide the devices into disjoint groups $\{\mathcal{D}_1, \mathcal{D}_2, \cdots, \mathcal{D}_m\}$, where devices in a group are predicted to belong to the same user.  
 \end{definition}
\vspace{-2mm}

  \noindent {\bf Design goals:} 
{\bf 1) Leveraging both IPs and domains.} IPs and domains are complementary information sources. 
Specifically, IPs used by devices represent their geolocations, while visited domains could indicate interests of the users.  Therefore, when both IPs used by devices and domains visited by devices are available, our method should leverage both of them to enhance tracking performance. However, our method should also be applicable when only IPs or domains are available, e.g., a first-party tracker may only be able to collect IPs used by devices. 

{\bf 2) Capturing latent correlations among IPs/domains.} 
IPs used by devices (or domains visited by devices) could have latent correlations. 
For instance, both Facebook and Twitter are social media websites. 
Our method should be able to 
discover such latent correlations and leverage them to track devices.

{\bf 3) Without requiring labeled device pairs.} In some scenarios, the tracker may be able to obtain some labeled device pairs. For instance, a first-party tracker (e.g., Facebook) could use cross-device IDs to obtain labeled device pairs from the users who log into 
the tracker's web services on multiple devices.  A third-party tracker (e.g., \texttt{Facebook like button}) who also serves as the first party could also use cross-device IDs to obtain labeled device pairs from the users who visit the web services that embed the tracker (e.g., \texttt{Facebook like button}). However, in some other scenarios, it is challenging for the tracker to obtain labeled device pairs. For instance, a pure third-party tracker (e.g., \texttt{Atlas}) that does not serve as the first party can hardly obtain labeled device pairs.
Hence, our method should be applicable with or without labeled device pairs.

{\bf 4) Robust to uncertainty in single-device tracking.} Cross-device tracking relies on single-device tracking to reliably link the browsing histories to devices. However, single-device tracking often has uncertainty, e.g., an IP used by one device or a domain visited by one device is incorrectly linked to another device. 
For instance, the same IP may be used by different users/devices due to NAT; 
If cookie or super-cookie based single-device tracking is adopted, a user may clear web cookies, or switch to private browsing mode or another browser, so that browsing histories of the same device may be linked to two different ones.
Moreover, browser fingerprinting, a new-generation of single-device tracking widely adopted by top Alexa websites~\cite{fmeasure}, is unreliable itself. According to 
a recent large-scale study, i.e., amIUnique~\cite{laperdrix2016beauty}, the accuracy of browser fingerprinting is only 89.4\%. That is, two devices sharing the same fingerprinting and browsing histories of these two actual devices may be linked to one inferred 
device.
Therefore, our method should be robust to uncertainty in single-device tracking.

IPFootprint~\cite{cao2015recovering}   does not satisfy any of the four requirements, DeviceGraph~\cite{deviceGraphKDD17} does not satisfy requirements 1), 2), and 4), and BAT-SU~\cite{zimmeck2017privacy} does not satisfy 2), 3), and 4).

\vspace{-1mm}
\section{GraphTrack}
\label{model}

\subsection{Overview}
The first challenge for cross-device tracking is how to leverage heterogeneous data sources, i.e., IPs and domains, to identify the correlations between devices. 
We address the challenge by modeling the interplays between IPs, domains, and devices as graphs, and leveraging graph mining techniques (in particular \emph{random walk with restart} on graphs) to capture the similarities between devices. The second challenge is how to match different devices without manual labels, i.e., in an \emph{unsupervised} fashion. We address the challenge by leveraging the \emph{symmetry} between devices. Specifically, our GraphTrack predicts a group of devices to match only if each pair of them has a large similarity score which we compute via analyzing the graph structure.  
The third challenge is how to be robust to uncertainty in single-device tracking that links browsing histories to devices. 
We address the challenge by  using the frequency a device accessed/visited a certain IP/domain.   

Next, we first discuss GraphTrack for 
unsupervised cross-device tracking. Then, we discuss how to incorporate manual labels into GraphTrack if they are available. 
{Finally, we analyze the computational complexity of our unsupervised and supervised 
methods.} 

\begin{figure}[t]
\vspace{-3mm}
\centering
\subfloat[IP-Device graph]{\includegraphics[width=0.2\textwidth]{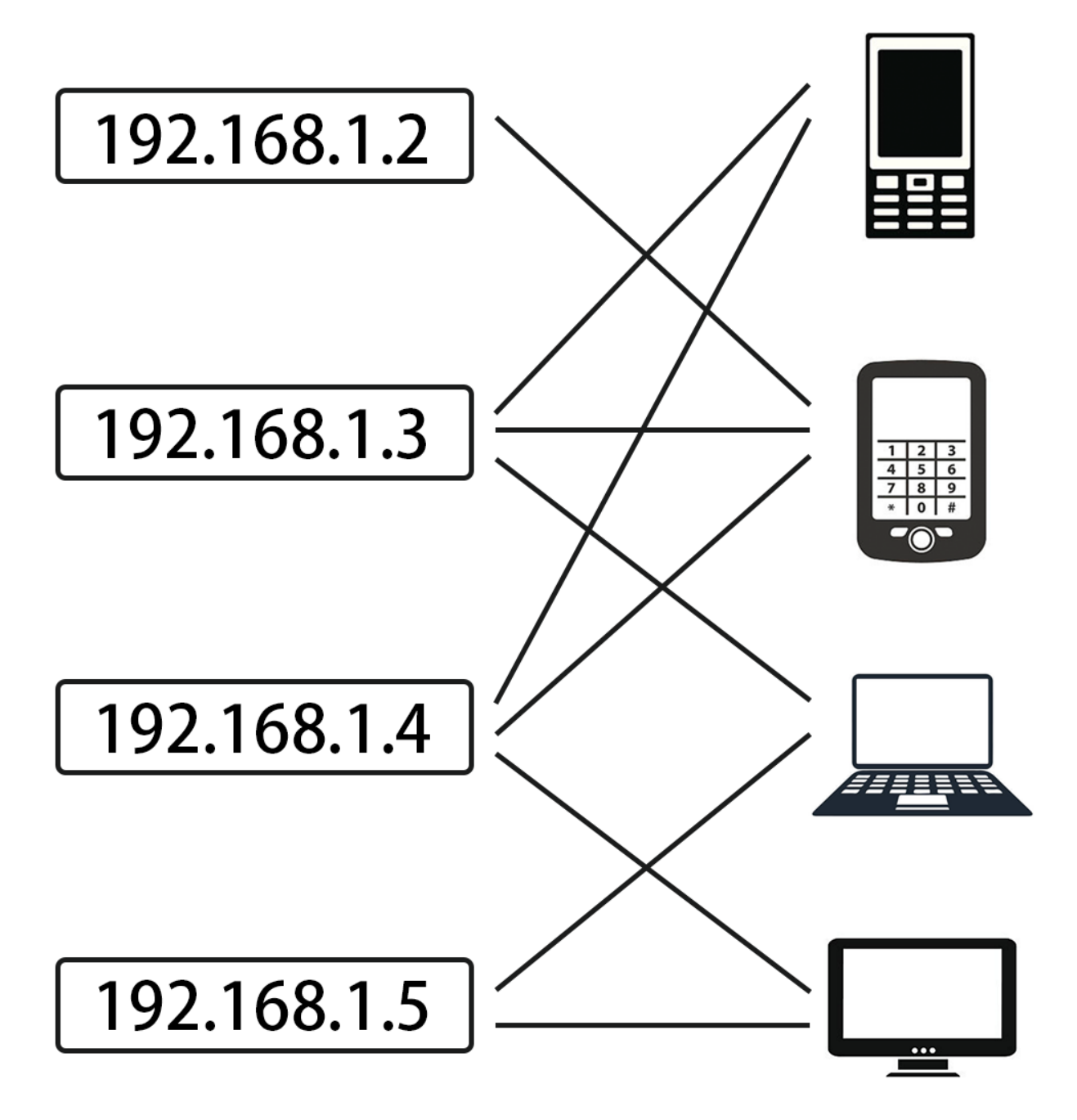}  \label{IPDeviceGraph}}
\subfloat[Domain-Device graph]{\includegraphics[width=0.2\textwidth]{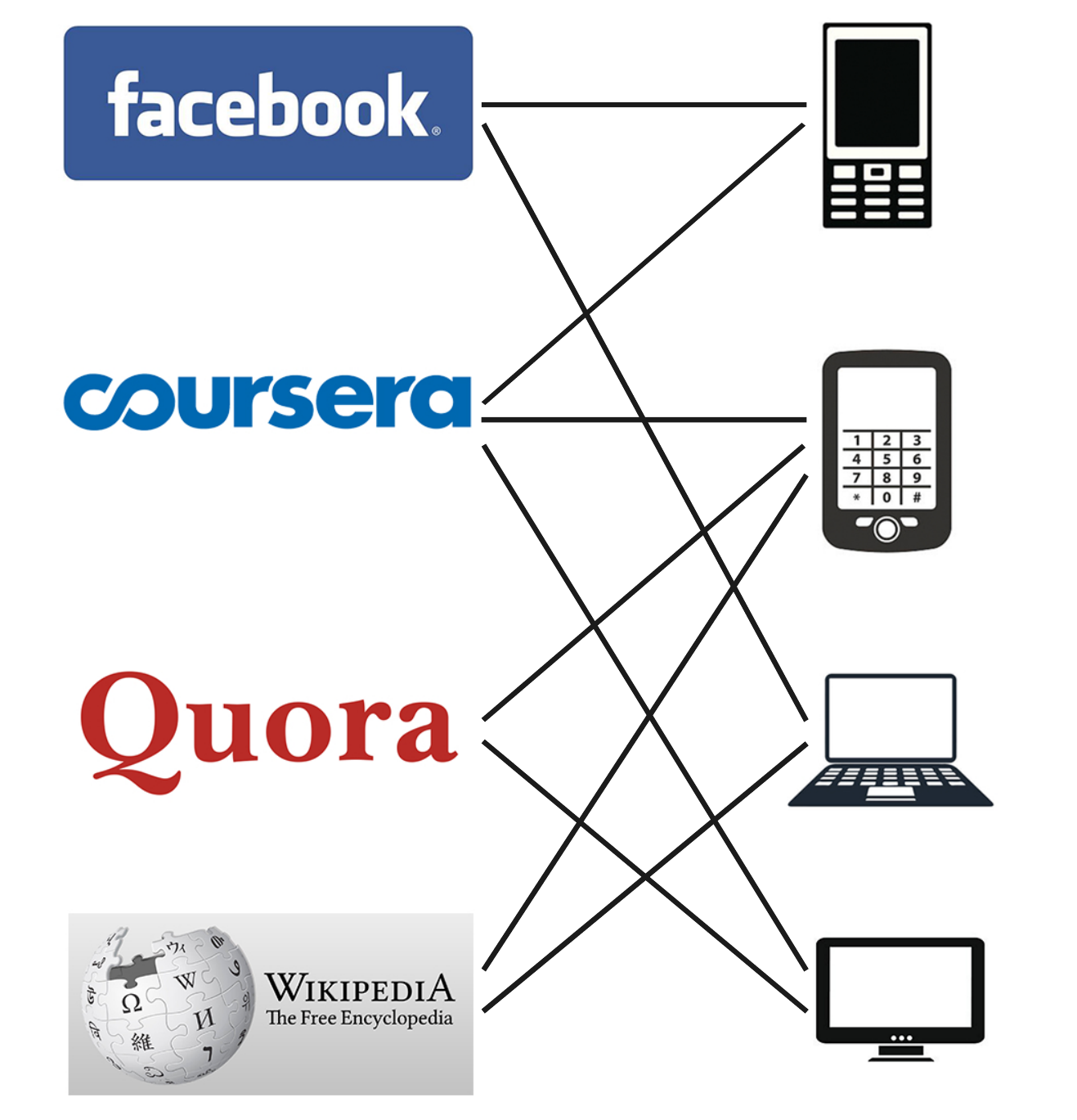} \label{DomainDeviceGraph} }
\vspace{-2mm}
\caption{(a) An example IP-Device graph. (b) An example Domain-Device graph. }
\vspace{-4mm}
\label{single-graph}
\end{figure}

\subsection{Unsupervised Cross-Device Tracking}
First, we model the interplays between IPs (or domains) and devices as graphs. Second, we adapt \emph{random walk with restart}, a popular graph mining technique, to our graphs to model similarities between devices. Third, we discuss  cross-device tracking using only IPs or domains based on the device similarities. 
Fourth, we combine IPs and domains for cross-device tracking.
\vspace{-2mm}
\subsubsection{Modeling IPs, Domains, and Devices via Graphs} We propose \emph{IP-Device graph} and \emph{Domain-Device graph} to model 
interplays between IPs, domains, and devices. Then, we use random walks with restart on the graphs to compute similarity scores between devices. 

\noindent {\bf Modeling interplays between IPs and devices as an IP-Device graph:} We represent each unique IP address (or IP prefix) and each device as a node. We create an edge between an IP node and a device node if the device used the IP at least once. Moreover, we model the edge weight of an edge between a device and an IP as the number of times that the device used the IP in its browsing history. 
We call this weighted graph \emph{IP-Device graph}. Figure~\ref{IPDeviceGraph} shows an example IP-Device graph. Note that there are no edges between device nodes. 
Our GraphTrack leverages the edge weights in the IP-Device graph to be more robust to uncertainty of single-device tracking at linking IPs to devices.
For instance, suppose a device used an IP multiple times in its browsing history; once a single-device tracker links a majority of them to the device, the corresponding edge still has a large weight. However, suppose a device did not use a certain IP, but the single-device tracker occasionally links the IP with the device. Then, the weight of the corresponding edge is small, and thus the incorrectly linked IP has a small impact on the overall structure of the IP-Device graph.

\noindent {\bf Modeling interplays between domains and devices as a Doma
in-Device graph:} 
We represent each unique domain and each device as a node; we create an edge between a domain node and a device node if the device visited the domain at least once; and 
we model the weight of an edge between a device and a domain as the number of times that the device visited the domain. We call this weighted graph \emph{Domain-Device graph}. Figure~\ref{DomainDeviceGraph} shows an example.

\vspace{-2mm}
\subsubsection{Modeling Device Similarity using Random Walk with Restart on Graphs} 
\label{randomwalk}

We leverage \emph{random walk with restart (RWwR)}~\cite{tong2006fast} 
(also known as \emph{Personalized PageRank}~\cite{pagerank}) 
to analyze the structure of the IP-Device graph and Domain-Device graph and model similarity between devices, {which can  capture semantic correlations among IPs and domains}. 
A larger similarity score indicates a higher likelihood of match. 
Next, we first introduce RWwR on a general weighted graph. 
Then, we adapt RWwR to compute similarity scores between devices in IP-Device and Domain-Device graphs. 

\noindent {\bf Random walk with restart (RWwR) on graphs:} 
Suppose we have an undirected weighted graph $G=(V, E, W)$, where 
$V$, $E$, and $W$ are the set of nodes, edges, and edge weights, respectively. 
For instance, $(u,v)$ is an edge between nodes $u$ and $v$, while $w_{u,v}$ is the weight of the edge $(u,v)$.  
We perform an RWwR in the graph from a seed node $u$. Specifically, in an RWwR, we have a particle that can stay on nodes in the graph. Initially, the particle is on node $u$. The particle iteratively moves along edges in the graph or jumps back to the initial node $u$. Suppose in the $t$th step, the particle is on node $v$. In the $(t+1)$th step, the particle jumps back to the initial node $u$ with a certain probability $\alpha$ (i.e., the random walk is restarted); and with the remaining probability $1-\alpha$, the particle picks a neighbor $x$ of $v$ with a probability proportional to the edge weight $w_{x,v}$ and moves to the neighbor. $\alpha$ is called \emph{restart probability}. Suppose $p_{u,v}$ is the frequency that the particle stays on node $v$. 
When the RWwR repeats for a large number of steps, $p_{u,v}$ becomes the probability that the particle will stay on node $v$ in each step. Conventionally, $\mathbf{p}_u = [p_{u,1}, p_{u,2}, \cdots, p_{u,|V|}]$ is called the \emph{stationary distribution} of the RWwR.
$p_{u,v}$ is a natural metric to measure similarity between $u$ and $v$. A larger $p_{u,v}$ indicates $v$ is structurally closer to $u$ on the graph and thus $v$ is more similar to $u$. Such RWwR based similarity was applied to rank relevant webpages in search engines~\cite{pagerank}, recommend accounts users wish to follow in Twitter~\cite{TwitterPageRank}, detect spammers in social networks~\cite{Yang12-spam,jia2017random,wang2018structure}, infer user attributes in social networks~\cite{gong2016you}, and infer social links from mobility profiles~\cite{backes2017walk2friends}, etc. 

Many techniques (e.g.,~\cite{tong2006fast,fastPageRank17}) have been developed to compute 
the stationary distribution $\mathbf{p}_u$ efficiently by theoretical computer science and data mining communities. 
For instance, $\mathbf{p}_u$ can be iteratively computed as follows:
{
\begin{align}
\label{computeP}
\mathbf{p}_u^{(t+1)} = (1 - \alpha) \mathbf{A} \mathbf{p}_u^{(t)}  + \alpha \mathbf{1}_u,
\end{align}
}%
where $\mathbf{p}_u^{(t)} = [p_{u,1}^{(t)}, p_{u,2}^{(t)}, \cdots, p_{u,|V|}^{(t)}]$ is the probability distribution of the RWwR in the $t$th iteration, $\mathbf{A}$ is the transition matrix of the graph with $|V|$ rows and $|V|$ columns, and $\mathbf{1}_u$ is an unit vector whose $u$th entry is 1 and other entries are 0. 
The $(u,v)$th entry of the transition matrix is formally 
defined as:
{
\begin{align}
\label{transition}
A_{u,v} = 
\begin{cases}
\frac{w_{u,v}}{d_v} &\text{ if } v\in \Gamma_u;  \\
0 &\text{ otherwise,} 
\end{cases}
\end{align}
}%
where $\Gamma_u$ is the set of neighbors of $u$ and $d_v$ is the weighted degree of $v$, i.e., $d_v=\sum_{s\in \Gamma_v} w_{s,v}$. To compute $\mathbf{p}_u$, we initialize a random vector $\mathbf{p}_u^{(0)}$ and then iteratively apply Equation~\ref{computeP} until the difference in two consecutive iterations is smaller than a certain threshold, e.g., $|\mathbf{p}_u^{(t+1)}-\mathbf{p}_u^{(t)}|_1 < 10^{-3}$.

\noindent {\bf Adapting RWwR to IP-Device and Domain-Device graphs to compute similarity scores between devices:} 
{State-of-the-art method~\cite{zimmeck2017privacy} computes the similarity scores between devices via Bhattacharyya coefficient, which is a \emph{simple} weighted common neighbor metric~\cite{murata2007link} and weights are the normalized frequencies of  IPs or domains. As a result, Bhattacharyya coefficient is unable to capture latent correlations among IPs and domains. 
In contrast, 
we adapt RWwR on graphs to model similarity between devices, which can capture latent correlations among IPs and domains.}
Specifically, we compute an \emph{IP-based similarity score} between two devices $D_i$ and $D_j$ (denoted as $s_{IP}(D_i, D_j)$) using an RWwR on the IP-Device graph; and we compute a \emph{domain-based similarity score} between two devices $D_i$ and $D_j$  (denoted as $s_{DO}(D_i, D_j)$) using an RWwR on the Domain-Device graph.

We take computing IP-based similarity scores as an example to illustrate more details. Suppose we have an IP-Device graph and a device $D_i$, we aim to compute similarity scores between $D_i$ and every device. One way is to simply apply an RWwR from $D_i$ in the weighted IP-Device graph, compute the stationary distribution $\mathbf{p}_{D_i}$ of the RWwR, and  define the IP-based similarity score $s_{IP}(D_i, D_j)$ between $D_i$ and $D_j$ as $s_{IP}(D_i, D_j)$ = ${p}_{D_i, D_j}$, for any $D_j \in \mathcal{D}$. However, as 
we will demonstrated in our experiments, such RWwR based similarity scores achieve suboptimal performance. This is because such similarity scores are heavily influenced by the weighted degree of a device, i.e., the total number of IP visits of a device. For instance, suppose a device $D_k$ belongs to a heavy Internet user and has a dense browsing history. Thus, in the IP-Device graph, many edges of $D_k$ have large weights. As a result, when the particle in the RWwR is on an IP node, the particle is more likely to move from the IP node to $D_k$ compared to other devices, which means $D_k$ will have a larger stationary probability and thus a larger similarity score with $D_i$. However, such similarity score is heavily biased by the weighted degrees of devices.

To address the issue, we propose to normalize the weight of an edge by the weighted degree of the corresponding device. 
Suppose an edge $(x,y)$ connects an IP $x$ and a device $y$.  We define a normalized edge weight $w_{x,y}'$ as 
$w_{x,y}'=\frac{w_{x,y}}{\sum_{z \in \Gamma_y} w_{z,y}}$. 
Then, we define the transition matrix $\mathbf{A}$ in Equation~\ref{transition} in an RWwR using the normalized edge weights.
In order to compute IP-based similarity scores between $D_i$ and every device, we perform an RWwR from $D_i$ in the IP-Device graph using $\mathbf{A}$. 
A larger similarity score $s_{IP}(D_i, D_j)$ between $D_i$ and $D_j$ indicates that $D_j$ is structurally closer to $D_i$ in the IP-Device graph and thus $D_j$ is more likely to match $D_i$.
Moreover, we leverage RWwR to compute similarity scores $s_{IP}(D_j, D_i)$ between $D_j$ and $D_i$. 
Note that 
$s_{IP}(D_i, D_j)$ and $s_{IP}(D_j, D_i)$ are 
very likely to be different because the random walks restart. 

Similarly, we compute 
$s_{DO}(D_i,$  $D_j)$ and $s_{DO}(D_j, D_i)$ using RWwR on the Domain-Device graph with the adapted transition matrix.

\begin{figure}[t]
\centering
\includegraphics[width=0.4\textwidth]{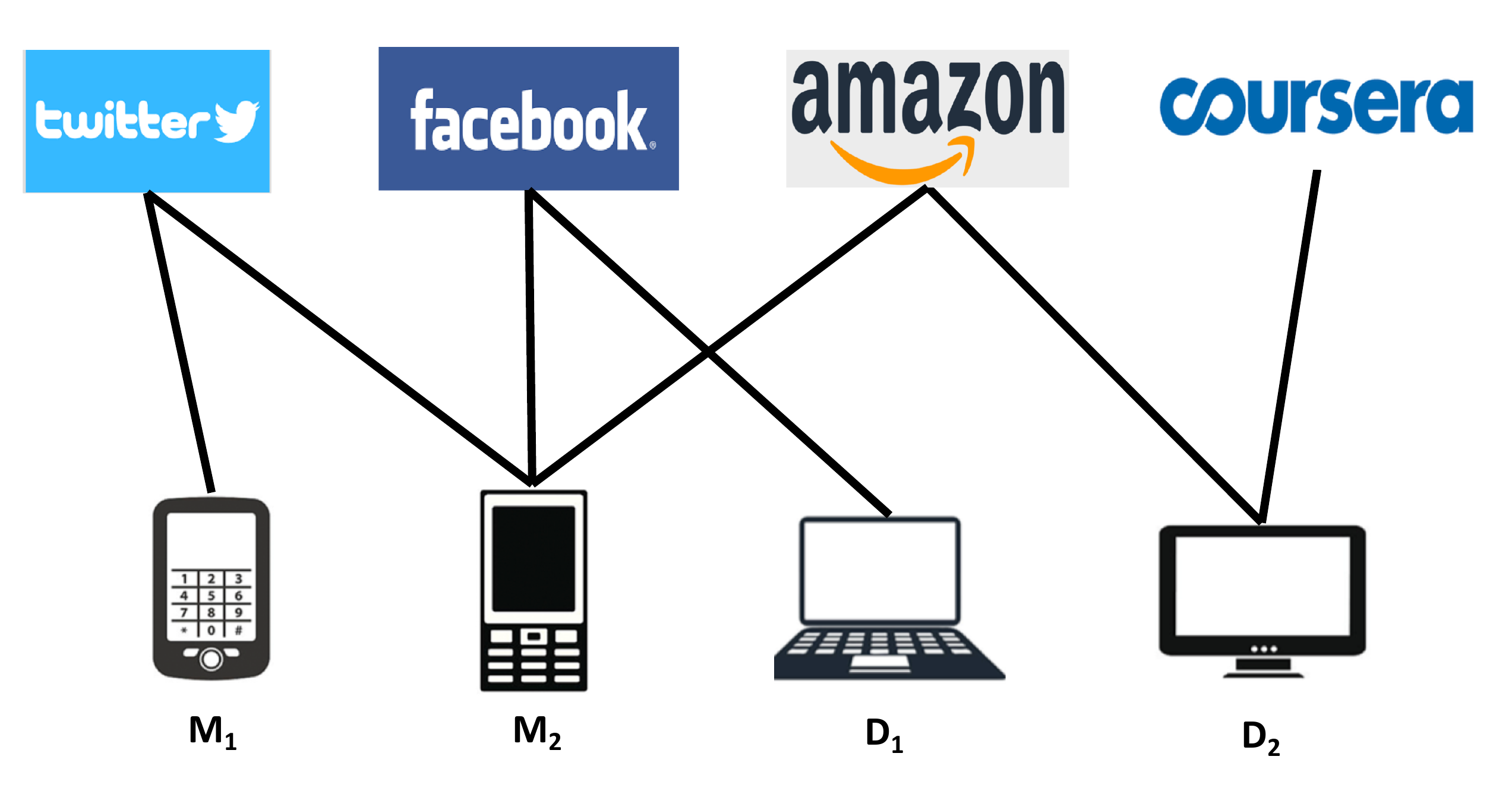}
\vspace{-2mm}
\caption{An example Domain-Device graph to illustrate how our adapted RWwR can capture latent correlations, while the state-of-the-art method cannot. 
}
\label{toyexample}
\vspace{-6mm}
\end{figure}

\vspace{-2mm}
\subsubsection{A Running Example} 
We use an example Domain-Device graph to tell the difference between the similarity scores computed by RWwR and those computed by the state-of-the-art BAT-SU~\cite{zimmeck2017privacy}.
Figure~\ref{toyexample} shows the example graph. 
In the example, a user has a mobile  $M_1$ and a desktop  $D_1$, i.e. $M_1$ matches $D_1$ as a ground truth. 
{The user visits Twitter on $M_1$ and Facebook on $D_1$. 
Two other devices $M_2$ and $D_2$ do not belong to the user.  
$M_2$ visits Facebook, Twitter, and Amazon, while $D_2$ visits Amazon and Coursera.
We assume that all devices visit these domains with the same frequency. 

 BAT-SU essentially computes the similarity score of two devices using the common neighbors on a Domain-Device graph. 
 Specifically, on the Domain-Device graph, for each common domain visited by the two devices,  BAT-SU computes a similarity score as the square root of the product of the normalized edge weights of the two corresponding edges. Then, BAT-SU adds such similarity scores for all common domains to get the final similarity score. In our example, BAT-SU computes the similarity score between the user's mobile device $M_1$ and desktop device $D_1$ as 0 since they do not have common neighbors. 
In RWwR, the similarity score between  $M_1$ and $D_1$ is non-zero. Specifically, to compute the similarity between $M_1$ and $D_1$, we start a random walk from $M_1$. The particle will move to the Twitter node in the next time step. Then, the particle could continue moving to $M_2$, the Facebook node, and $D_1$. Therefore, when the random walk converges, $D_1$ would have non-zero stationary probability. Moreover, we calculate that $s_{DO}(M_1, D_1) = 0.032 > s_{DO}(M_1, D_2) = 0.026$, which indicates that $M_1$ correctly matches $D_1$. 
The reason is that RWwR can capture the latent relationships between Twitter and Facebook via using the data from other users, who visit both Facebook and Twitter.} 

{We note that such latent correlations do exist in practice. In particular, in the 
dataset~\cite{zimmeck2017privacy} we use in Section~\ref{Eval}, domains are anonymized by hashing. We found user 49 did not visit domain A 
(hash value: {\small b17e6ac8f4740bb465c6acf5cd0052b9bbc5ef49a0e58569eb00a87e
2831ac5f}) 
but frequently visited domain B (hash value: {\small 2da7c187647dc68
9d1686d250b24aebb3f551b24f88d90e58c2d49d5a3f1c617}) on its mobile device. In particular, the user visited domain B 169 times on its mobile device, which is among the top-5 most frequently visited domains on its mobile device. However, the user frequently visited domain A  but did not visit domain B on its desktop device. In particular, the user visited domain A 174 times, which is among the top-5 most frequently visited domains on its desktop device. We also found that 162 devices frequently visited both domains (363 and 862 times on average, respectively). 
Overall, we found that 8 out of the 107 users are such users in the dataset. 
Our GraphTrack can correctly match the mobile device and the desktop device of all the 8 users, while BAT-SU cannot correctly match any of them.}

\subsubsection{GraphTrack using IPs or Domains Alone}
\label{framework}

Suppose we have $n$ devices $\mathcal{D}=\{D_1, D_2, \cdots, D_n\}$. 
For each device, we have a list of IPs used by the device and domains visited by the device in its browsing history. 
We assume each user has at most $K$ devices. 
Intuitively, if two devices belong to the same user, then one device is  likely to have a top-$(K-1)$ largest similarity score with the other device among all devices and vice versa. 
Our GraphTrack leverages such \emph{symmetry} between similarity to match devices. 
Next, we use IPs to illustrate the details of our GraphTrack. Specifically, GraphTrack has the following three steps.  

\begin{itemize}[leftmargin=*]
\item \myparatight{Step I} We perform an RWwR from each device $D_i$ in the IP-Device graph with the adapted transition matrix to compute similarity scores between $D_i$ and every device. 
Then, we find the $K-1$ devices that have the largest similarity scores with $D_i$ and we denote them as a set $\mathcal{D}_{i,IP}$.

\item \myparatight{Step II} We construct a device \emph{similarity graph}, where each node represents a device and we add an edge  between two devices  if and only if they are among each other's top-$(K-1)$ list of most similar devices. Specifically, we add an edge between two devices  $D_i$ and $D_j$ if and only if $D_i \in \mathcal{D}_{j,IP}$ and $D_j \in \mathcal{D}_{i,IP}$.

\item \myparatight{Step III} Finally,  GraphTrack predicts the devices in a \emph{clique} in the similarity graph belong to the same user. A clique is a subgraph where each pair of nodes are connected {directly}.  Note that a device may appear in multiple cliques. If this happens, we predict the device belongs to the largest clique (to track users that have a large number of devices) or randomly assign it to a clique if the cliques have the same size. If a device forms a clique by itself,  GraphTrack predicts the device has no match.
\end{itemize}

Similarly, we can use domains alone to perform tracking. We denote GraphTrack that use IPs and domains alone as \emph{GraphTrack-IP} and \emph{GraphTrack-Domain}, respectively.

\subsubsection{Combining IPs and Domains}
\label{combine}
We propose three methods to combine IPs and domains.

\noindent {\bf GraphTrack-UniGraph:} 
It integrates IPs, domains, and devices in a single unified graph, which we call \emph{IP-Device-Domain graph}. 
Specifically, in the IP-Device-Domain graph, we represent each unique IP, unique domain, and each device as a node; and we create an edge between an IP (or domain) and a device if the device used the IP (or visited the domain). 
Then, we construct a device similarity graph based on the IP-Device-Domain graph, similar to what we did on the IP-Device graph.
Finally, we perform clique analysis in the similarity graph to identify matched devices.   
We denote this variant of GraphTrack as \emph{GraphTrack-UniGraph}.

\noindent {\bf GraphTrack-OR:} 
It combines GraphTrack-IP and GraphTrack-Domain via the \emph{OR} operator. 
Specifically, GraphTrack-OR computes a similarity graph based on the similarity graphs computed by GraphTrack-IP and GraphTrack-Domain. In particular, two devices are connected in the similarity graph if they are connected in the similarity graph constructed by either GraphTrack-IP or GraphTrack-Domain. Then, we perform clique analysis in the similarity graph to identify matched devices.

\noindent {\bf GraphTrack-AND:} 
It combines GraphTrack-IP and GraphTrack-Domain via the \emph{AND} operator.  
Specifically, we create a similarity graph, where two devices are connected if they are connected in the similarity graphs constructed by both GraphTrack-IP and GraphTrack-Domain. Then, we use clique analysis to match devices.  

\subsection{Incorporating Manual Labels}
\label{graphtrack_sup}
We have five unsupervised variants of GraphTrack, i.e., GraphTrack-IP, GraphTrack-Domain, GraphTrack-UniGraph, GraphTrack-OR, and GraphTrack-AND. We discuss how to adapt them to incorporate manual labels if they are available. We append a suffix ``-SU'' to a method to indicate the supervised version, e.g., GraphTrack-IP-SU is the supervised version of GraphTrack-IP. 

\noindent {\bf GraphTrack-IP-SU, GraphTrack-Domain-SU, and GraphTrack-UniGraph-SU:} 
Suppose we have $L$ manually labeled device pairs. 
We denote the $L$ labeled pairs as $\mathcal{L} = \{(D_1, M_1), \cdots, (D_L, M_L)\}$, where $D_i, M_i \in \mathcal{D}, i=1, \cdots, L$.
Note that multiple labeled device pairs could belong to the same user. 
We use GraphTrack-IP-SU as an example to illustrate how we adapt GraphTrack to incorporate manual labels.  
 Our key idea is to learn a \emph{threshold} of similarity score using the labeled device pairs to determine whether two devices should be connected or not in the similarity graph. Specifically, we connect two devices in the similarity graph if one device is among the other device's top-$(K-1)$ list of most similar devices and their similarity score is no less than the threshold. Then, we leverage clique analysis to match devices.

Next, we discuss how to learn the threshold. Roughly speaking, we set the threshold as the smallest similarity score among the labeled device pairs. Specifically, we initialize a similarity score set $S=\emptyset$. 
For each labeled device pairs $(D_l, M_l)$, we compute their similarity scores $s_{IP}(D_l, M_l)$ and $s_{IP}(M_l, D_l)$. If  $M_l$ (or $D_l$) is among $D_l$'s (or $M_l$'s)  top-$(K-1)$ list of most similar devices, then we add $s_{IP}(D_l, M_l)$ (or $s_{IP}(M_l, D_l)$) into the set $S$.  
In the end, we set the threshold to be the minimum similarity score in the set $S$.

We adapt GraphTrack-Domain and GraphTrack-UniGraph to GraphTrack-Domain-SU and GraphTrack-UniGraph-SU in the same way, i.e., replacing the similarity scores in GraphTrack-IP-SU as those computed using the Domain-Device graph/IP-Device-Domain graph, respectively.

\noindent \textbf{GraphTrack-OR-SU and GraphTrack-AND-SU:} GraphTrack-OR-SU combines GraphTrack-IP-SU and GraphTrack-Domain-SU via the OR operator, while GraphTrack-AND-SU combines them via the AND operator. 
In the OR operator,  two devices are connected in the similarity graph if they are connected in the similarity graphs constructed by either  GraphTrack-IP-SU or GraphTrack-Domain-SU. In the AND operator, two devices are connected if they are connected in the similarity graphs constructed by both  GraphTrack-IP-SU and GraphTrack-Domain-SU.

\vspace{-2mm}
\subsection{Computational Complexity}
Due to the limited space, we defer the detailed computational complexity analysis to Appendix~\ref{comp}.

\section{Evaluation}
\label{Eval}

\subsection{Experimental Setup}
\label{expsetup}

\noindent {\bf Dataset description:}
We use two real-world datasets to perform cross-device tracking. One 
dataset 
is for mobile-desktop tracking, where all users have one mobile 
device 
and one desktop device; the other one
is for multiple-device tracking, where users have 2-5 devices. 
In most of our experiments, we focus on the mobile-desktop tracking dataset because it was used by previous method~\cite{zimmeck2017privacy} and has both IPs and domains. 

{\bf Mobile-desktop tracking dataset.} 
It was collected at Columbia University and is publicly available~\cite{zimmeck2017privacy}. The dataset contains 126 users, 
where 107 users have both a mobile device and a desktop device, and the remaining users have either a mobile device or a desktop device. The dataset includes the \emph{IP addresses} used by each device and \emph{Internet domains} visited by each device within $\sim$3 weeks. Like the previous study~\cite{zimmeck2017privacy}, we focus on the 107 users that have both a mobile device and a desktop device. The total number of unique IPs and domains are 7,290 and 17,140, respectively. Moreover, each device used 85 unique IPs and visited 315 unique domains on average. We note that the dataset also includes mobile apps used by each mobile device. However, such information has negligible impact on device tracking as shown by Zimmeck et al.~\cite{zimmeck2017privacy}. Thus, we will not consider such information in this work for simplicity. 
For mobile-desktop tracking, our task is to identify each user's desktop device given his mobile device, similar to~\cite{zimmeck2017privacy}.

On the mobile-desktop tracking dataset, we remove the top-50 domains that have the most visits, because these 50 domains occupy around 70\% of all visits, dominate the device similarities, and negatively impact the tracking performance. For instance, Figure~\ref{result-Domain-All} 
in the Appendix 
shows the performance of GraphTrack-Domain with and without the top-50 domains. GraphTrack-Domain achieves much better performance when filtering the top-50 domains.
We note that Zimmeck et al.~\cite{zimmeck2017privacy} filtered the top-50 domains ranked by Alexa and all columbia.edu domains. 
As the released dataset anonymized each domain by cryptographic hashing, we cannot apply the same filtering.

{\bf Multiple-device tracking dataset.} 
We collected this dataset between December 2017 and July 2018, in cooperation with a real-world website with regular users.\footnote{Due to confidentiality agreement, we cannot disclose the website name.} 
 {Each user, if included in the dataset, made a consent on the collected information during registration and our research was approved by IRB.} 
 Particularly, the website, acting as a first-party tracker, collects the following information related of a user device: the device's IP, the device's OS, 
the cookie ID for the website, 
and user ID (anonymized). 
We treat each ``device's OS + user ID" as a device. 
As the number of IPs is large, 
we select the IP prefix (i.e., the first 24 bits of an IP address) to represent each IP address.   
In the dataset, we have 173,279 unique IP prefixes, 335,434 unique devices, 
and 154,508 users. 87.9\% of users have 2 devices, 10.5\% of users have 3 devices, 1.3\% of users have 4 devices, and 0.34\% of users have 5 devices. {Among the 173,279 unique IP prefixes, 22\% of them are accessed by devices from different users.} Note that, since the dataset was collected from a single website/domain, we only construct the IP-Device graph and use GraphTrack-IP and GraphTrack-IP-SU to perform cross-device tracking. 
Moreover, we found that using IP prefix achieves better performance than using the raw IP. For instance, when using 50\% of users' devices for training, GraphTrack-IP using the raw IP has an Accuracy 0.49, while GraphTrack-IP using the IP prefix has an Accuracy 0.60, where Accuracy is defined in the next paragraph.

\myparatight{Evaluation metrics} Like the previous study~\cite{zimmeck2017privacy}, we use 
\emph{Accuracy}, \emph{Precision}, \emph{Recall}, and \emph{F-Score} to evaluate cross-device tracking methods.  These metrics involve True Positives (TP), False Positives (FP), True Negatives (TN), and False Negatives (FN). Suppose a method predicts some matched device groups. 
TP is the number of groups that truly match, 
and FP is the number of groups that do not match in the groundtruth. 
Suppose a method predicts that some devices have no match.  TN is the number of such devices that truly have no matches, while FN is the number of such devices that actually have matches. 
Given these terms, we have  
\emph{Accuracy}=$\frac{TP+TN}{TP+TN+FP+FN}$, \emph{Precision}=$\frac{TP}{TP+FP}$, \emph{Recall}=$\frac{TP}{TP+FN}$, and \emph{F-Score} =$\frac{2\ *\ \emph{Precision}\ *\ \emph{Recall}}{\emph{Precision}+\emph{Recall}}$.

\myparatight{Compared methods}
We compare GraphTrack with supervised and unsupervised cross-device tracking methods. Section~\ref{app:comp} and
Table~\ref{eval_set} in Appendix summarizes all compared methods. 

\noindent {\bf Parameter setting:} We set the restart probability $\alpha$ in a random walk to be 0.15 as previous work~\cite{tong2006fast}. We also explored the impact of $\alpha$, but we found that it has a very small impact on 
the performance of GraphTrack.  
Moreover, {similar to previous work~\cite{tong2006fast}}, we set the number of iterations to be $\log|V|$, where $|V|$ is the number of nodes in a graph (IP-Device, Domain-Device, or IP-Device-Domain graph). 
We set $K=2$ for mobile-desktop tracking, as one mobile device has only one matched desktop device. 
We set $K=5$ for multiple-device tracking, as each user has at most 5 devices.  
For all unsupervised methods, we perform mobile-desktop tracking for each mobile device, i.e., for each mobile device in the dataset, we either predict the matched desktop device or predict no match; and we perform multiple-device tracking for all devices, i.e., for each device in the dataset, we predict all its matched devices or predict no match.
For supervised methods, we sample some users and use their device pairs as a training set and use the remaining users as the testing set.  
We repeat the experiment 5 times and report the mean and standard deviation. 
All experiments are conducted on a Linux machine with 512GB memory and 32 cores.

\begin{figure}[!t]
\vspace{-4mm}
\centering
\subfloat[]{\includegraphics[width=0.24\textwidth]{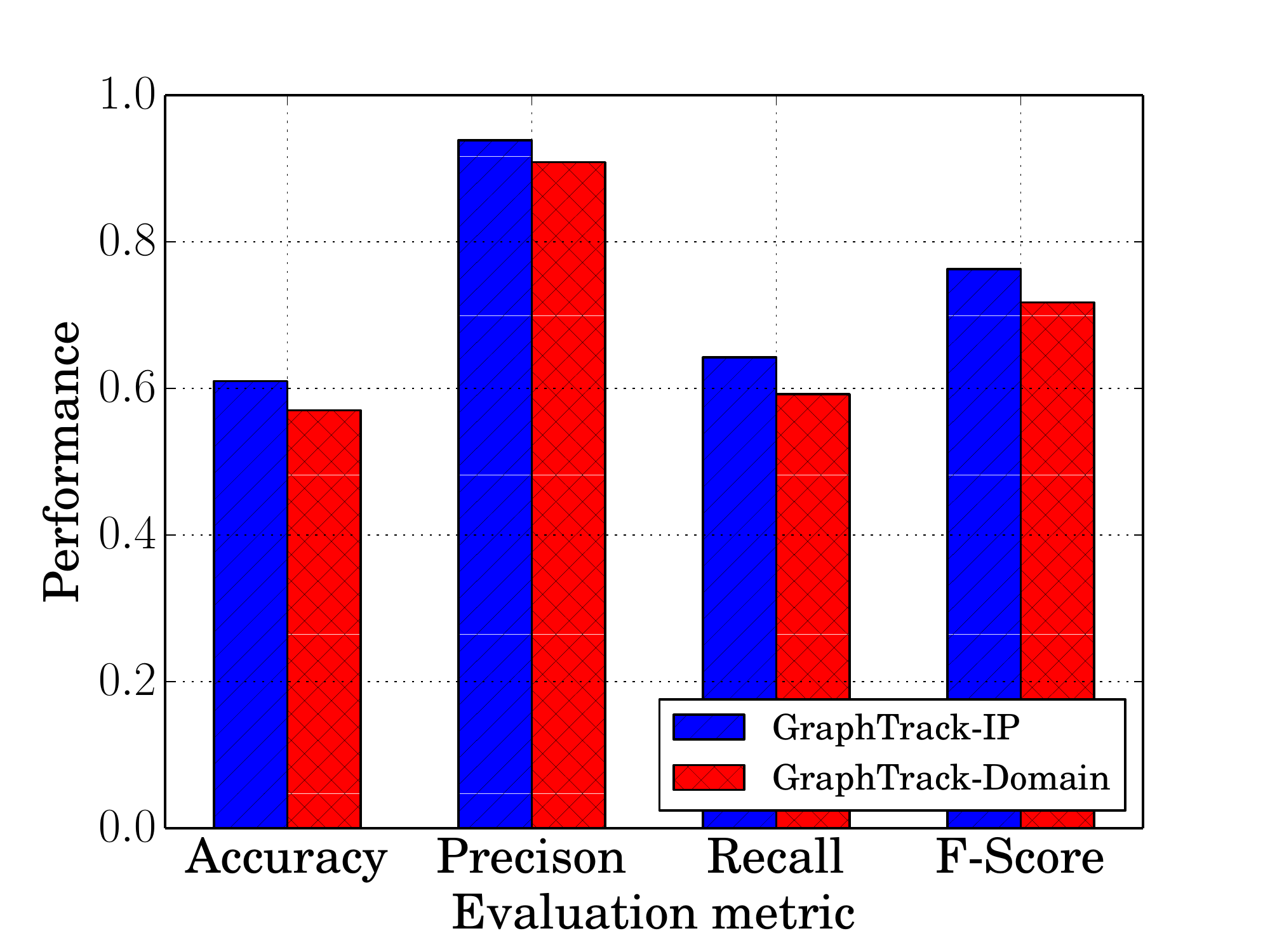} \label{result-single}}
\subfloat[]{\includegraphics[width=0.24\textwidth]{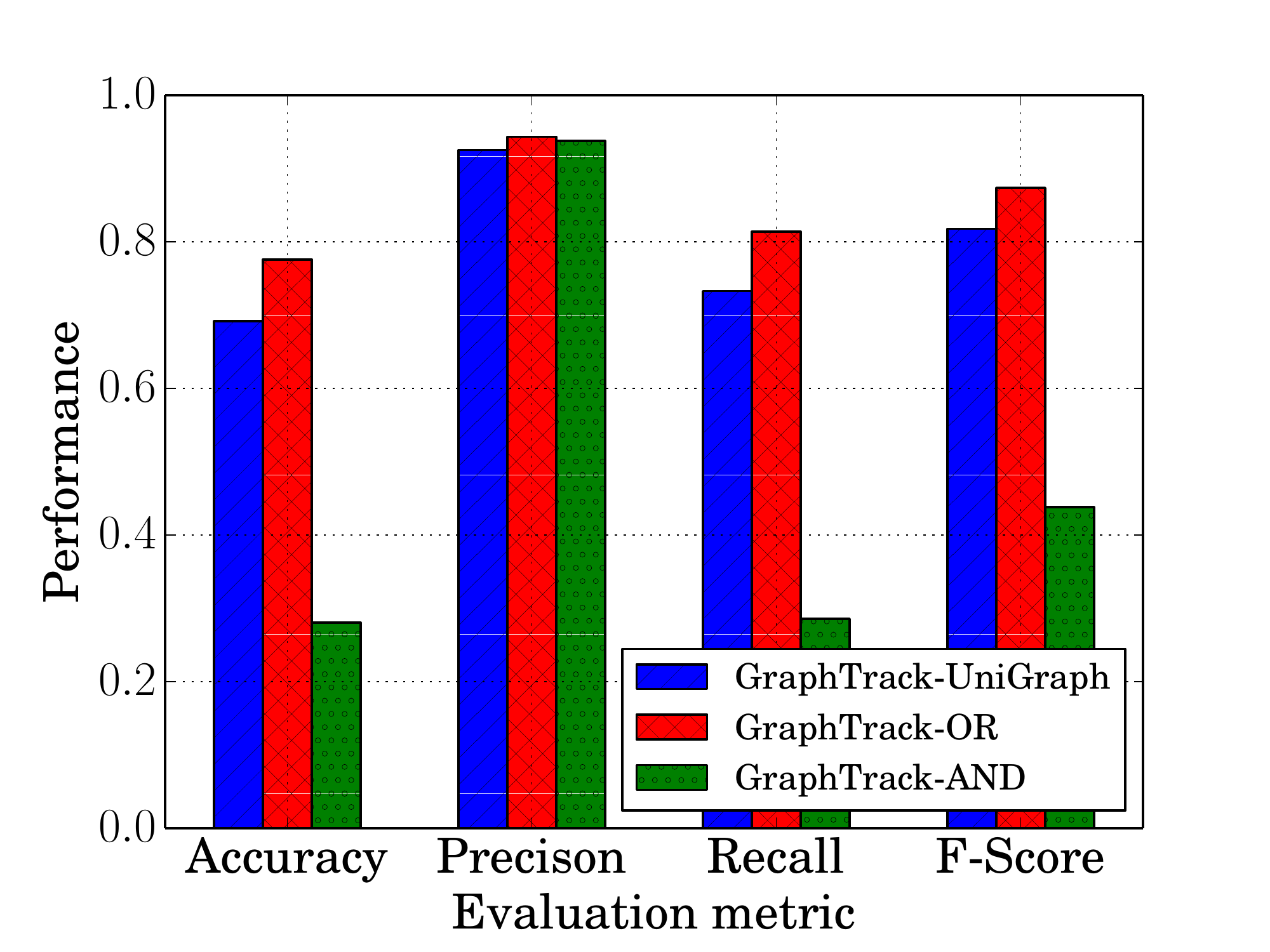} \label{result-combination}}
\vspace{-2mm}
\caption{(a) Comparing mobile-desktop tracking methods using IPs or domains alone. (b) Comparing different ways to combine IPs and domains.} 
\vspace{-6mm}
\end{figure}

\subsection{Mobile-Desktop Tracking Results for Unsupervised Methods}

In this part, we evaluate unsupervised GraphTrack, which would be
used in the scenario of a pure third-party tracker, such as ads embedded in first-party domains.  As users do not interact with these trackers, e.g., logging into the tracker's domain, such trackers have no access to labeled device pairs. 
Unless otherwise mentioned, we use GraphTrack-OR as it outperforms other versions of GraphTrack. 

\myparatight{IPs are more informative than domains} Since we perform mobile-desktop tracking using IPs and domains, one natural question is which type of data is more informative. 
To answer this question, we compare the performance of our methods GraphTrack-IP and GraphTrack-Domain. 
Figure~\ref{result-single} shows the results of GraphTrack-IP and GraphTrack-Domain at matching the mobile devices to desktops in the dataset. 
We observe that GraphTrack-IP performs slightly better than GraphTrack-Domain, which means that IPs are more informative than domains at tracking devices when using our GraphTrack method. 
{Two possible reasons are: 1) domains are more diverse than IPs on average in a device's browsing history; 2) IPs somewhat indicate location information of users but different users may visit similar domains. Thus, IPs can better distinguish between devices of different users.}
We note that previous studies~\cite{cao2015recovering, zimmeck2017privacy} also found that IPs are more informative than domains at cross-device tracking, which is consistent with our observation.

\begin{figure}[!t]
\vspace{-4mm}
\centering
\subfloat[]{\includegraphics[width=0.24\textwidth]{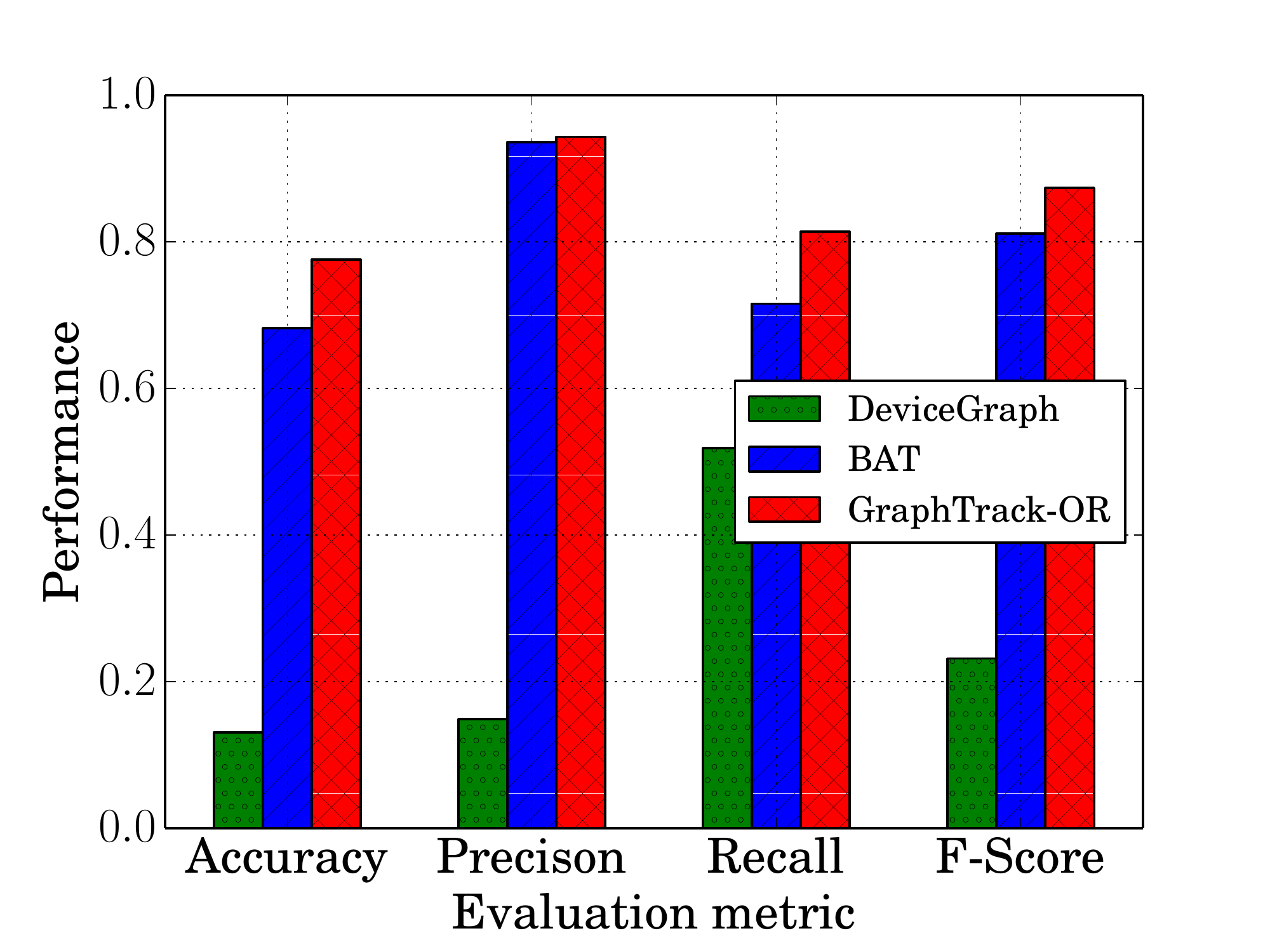} \label{result-unsup}}
\subfloat[]{\includegraphics[width=0.24\textwidth]{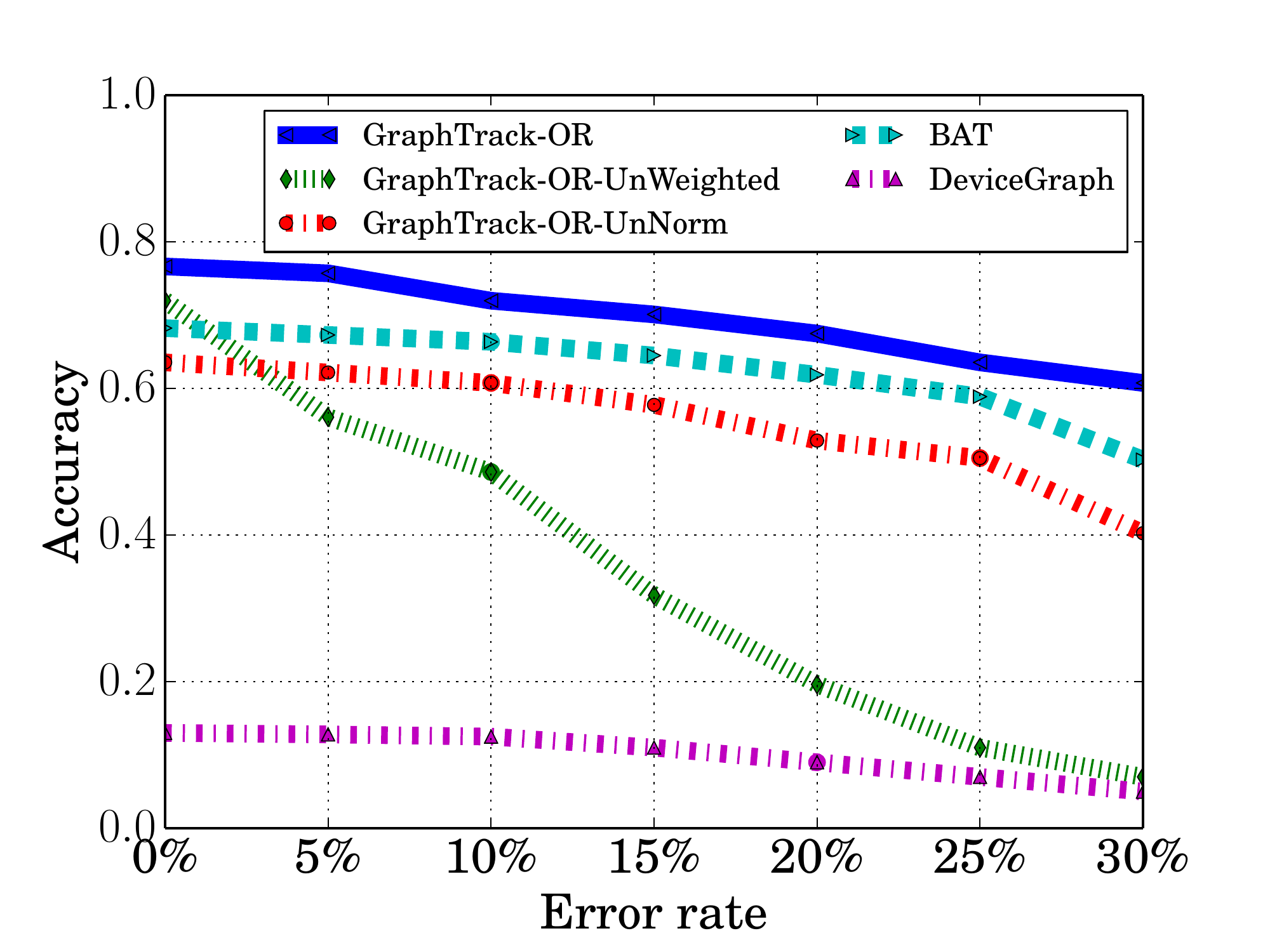} \label{result-noise}}
\vspace{-2mm}
\caption{(a) Comparing unsupervised cross-device tracking methods. (b) Impact of error rates in single-device tracking on Accuracy of cross-device tracking.} 
\vspace{-6mm}
\end{figure}

\noindent {\bf Comparing different ways to combine IPs and domains:} 
Figure~\ref{result-combination}
shows the comparison results among GraphTrack-UniGraph, GraphTrack-OR, and GraphTrack-AND. 
We see that GraphTrack-OR consistently performs better than GraphTrack-UniGraph and GraphTrack-AND. Therefore, in the rest of this section, we will focus on GraphTrack-OR.
GraphTrack-UniGraph combines IPs and domains using the IP-Device-Domain graph. This graph does not well distinguish the heterogeneous data types, i.e., IPs and domains. For instance, this graph does not distinguish between edges linking to IPs and edges linking to domains.  As a result, GraphTrack-UniGraph achieves inferior performance. Intuitively, GraphTrack-AND should have a high Precision because it has a ``higher'' standard to predict a match for a mobile-desktop pair. Specifically, a mobile device and a desktop device are predicted to match only if both GraphTrack-IP and GraphTrack-Domain predict a match. This explains why GraphTrack-AND has a higher Precision than GraphTrack-UniGraph, though GraphTrack-AND has a much lower Accuracy, Recall, and F-Score.
However, we found GraphTrack-AND achieves a lower Precision than GraphTrack-OR. The reason is, compared to GraphTrack-OR, GraphTrack-AND predicts a smaller number of both TPs and FPs, but its TPs is much smaller than FPs. 
\emph{Thus, in the rest of this section, we focus on GraphTrack-OR.}

\noindent {\bf GraphTrack outperforms compared unsupervised methods:}
Figure~\ref{result-unsup} shows the results of GraphTrack-OR, BAT, and DeviceGraph. We observe that our GraphTrack consistently outperforms BAT and significantly outperforms DeviceGraph. 
We note that BAT predicts more mobile devices to have no matches, i.e., BAT has more FNs than GraphTrack-OR.
The reason is that BAT essentially computes the common neighbors between a mobile device and a desktop device in the IP-Device graph and Domain-Device graph, while GraphTrack leverages RWwR to analyze the complex graph structure. 
DeviceGraph has bad performance because it predicts that a large number of unmatched devices fall into the same community, resulting in large FPs (i.e., 80 FPs in our experiments).

\begin{figure}[!tbp]
\vspace{-4mm}
\centering
\subfloat[]{\includegraphics[width=0.24\textwidth]{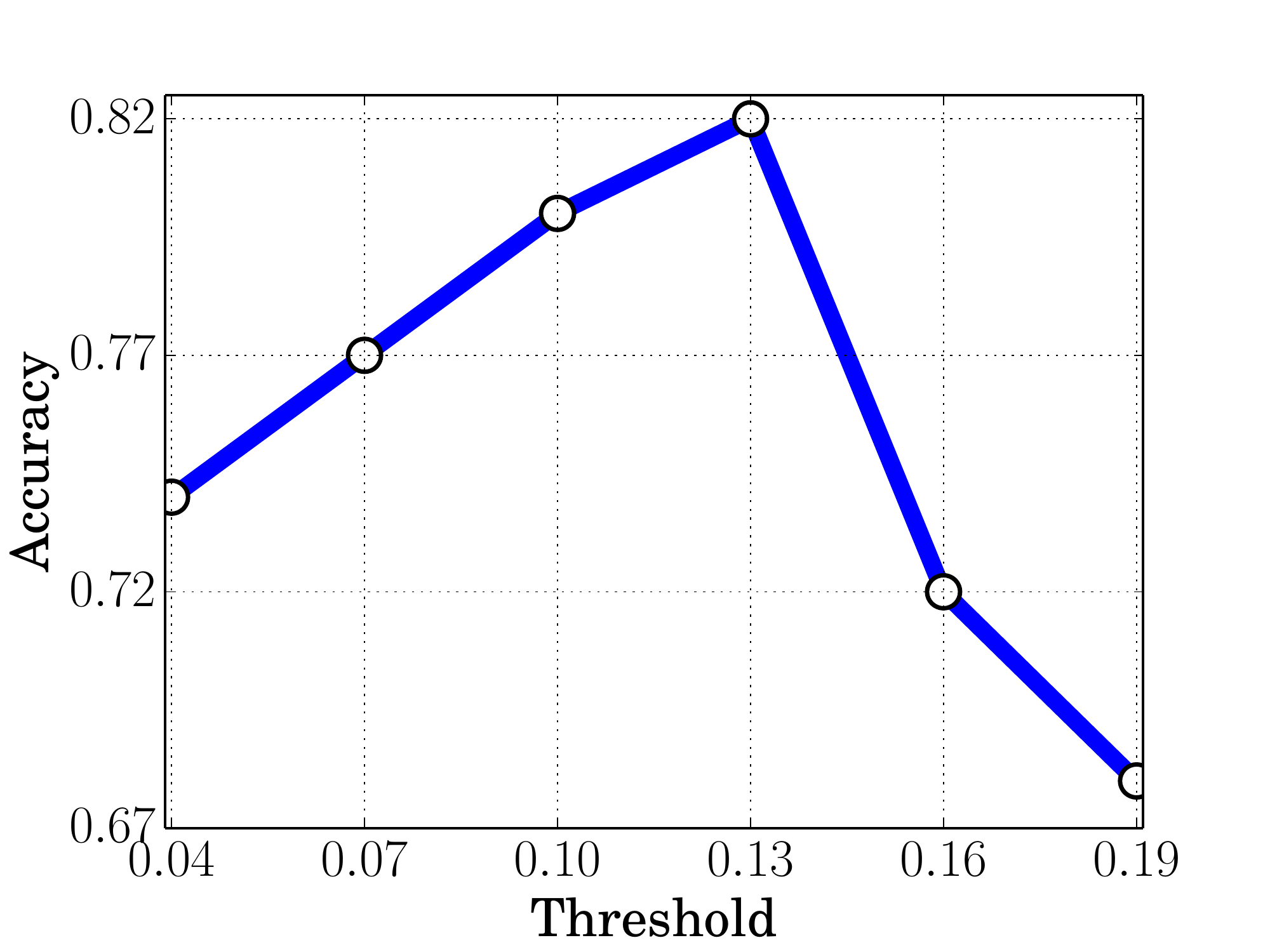}  \label{fig:threshold}}
\subfloat[]{\includegraphics[width=0.24\textwidth]{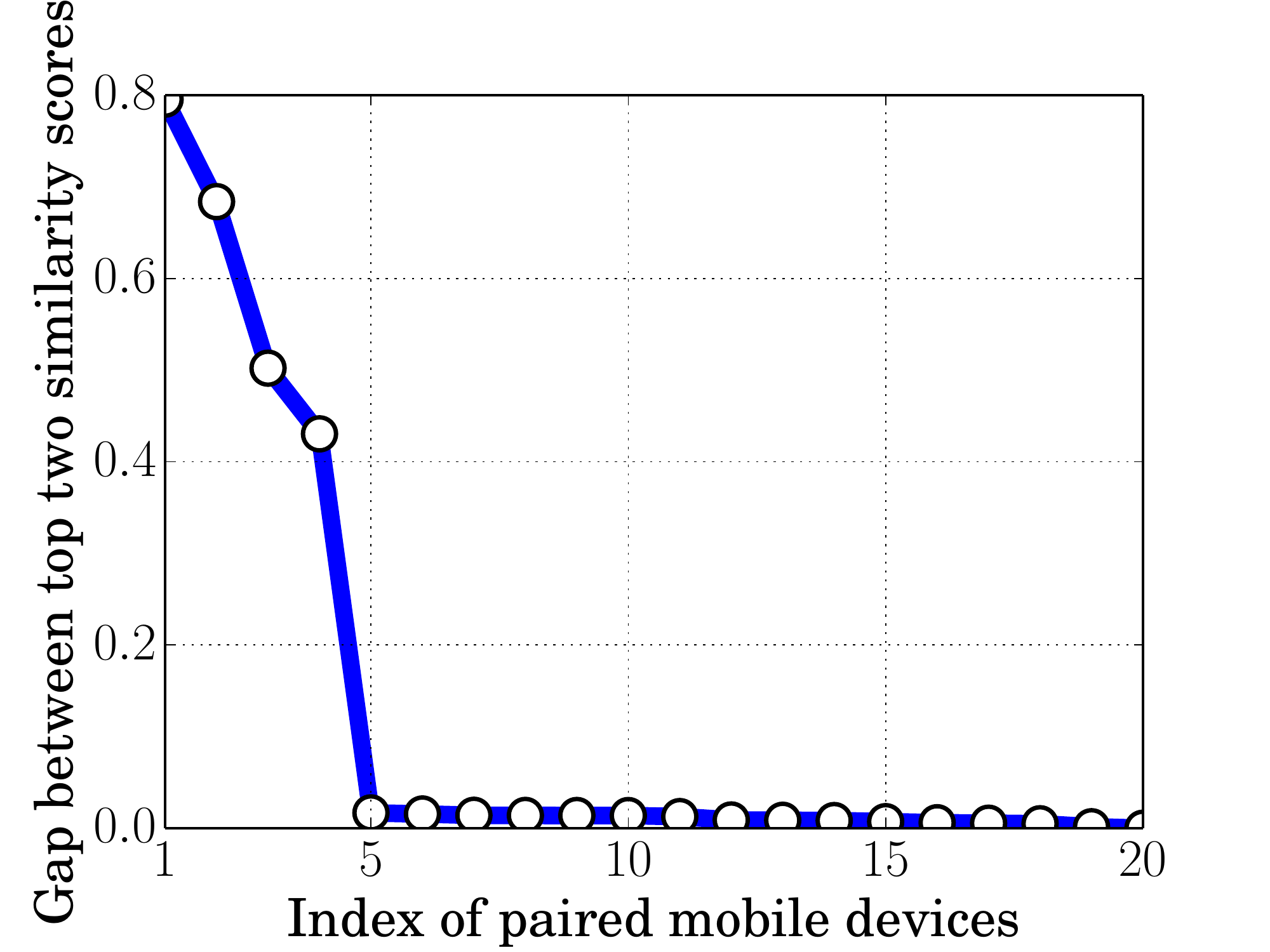} \label{fig:gap} }
\vspace{-2mm}
\caption{Alternatives of converting BAT-SU to be unsupervised. (a) Accuracy vs. threshold. (b) Gaps between top two similarity scores for 20 sampled mobile devices. }
\vspace{-6mm}
\label{sup-unsup}
\end{figure}

\noindent {\bf Robustness to uncertainty in single-device tracking:} 
Single-device tracking could have uncertainty, e.g., an IP or domain visited by one device may be linked to another device that did not access it. Thus, one natural question is how uncertainty in single-device tracking impacts the performance of cross-device tracking. 
Since our dataset does not allow us to implement a real-world single-device tracker, we simulate an inaccurate single-device tracker and study its impact on cross-device tracking. 
Specifically, for each web visit from a device, the single-device tracker incorrectly links the visit to a randomly selected wrong device with a certain probability $y\%$. We call $y\%$ the \emph{error rate} of the single-device tracker. 
To simulate an $y\%$ error rate,
we randomly sample $y\%$ of the web visits of each mobile (or desktop) device and randomly distribute them to other mobile (or desktop) devices. We then use the noisy web visits of each device to perform cross-device tracking.

Figure~\ref{result-noise} shows Accuracy of several mobile-desktop tracking methods as a function of error rates in single-device tracking. 
Recall that we discussed several ways to deal with edge weights in the IP-Device graph and Domain-Device graph in Section~\ref{model}.  
For instance, \emph{GraphTrack-OR-UnWeighted} indicates that we do not use edge weights in GraphTrack-OR, i.e., we set all edge weights to be 1. \emph{GraphTrack-OR-UnNorm} indicates that we do not normalize edge weights by the weighted degree of devices, i.e., the weight of an edge between a device and an IP (or domain) is the number of times the device visited the IP (or domain). 
We have several observations: First, GraphTrack-OR consistently outperforms BAT as the error rate increases. Second, GraphTrack-OR decreases slowly as the error rate increases, e.g., GraphTrack-OR's Accuracy only decreases by 0.04 when the error rate is 10\%. Third, GraphTrack-OR significantly outperforms GraphTrack-OR-UnWeighted and GraphTrack-OR-UnNorm. Specifically, when single-device tracking has no errors, GraphTrack-OR-UnWeighted has close performance with GraphTrack-OR. However, GraphTrack-OR-UnWeighted's performance quickly decreases as the error rate increases. This is because as the error rate increases, the IP-Device graph and the Domain-Device graph have much more edges that are noises, which significantly impact the graph structure based similarity scores.  
GraphTrack-OR-UnNorm is worse than GraphTrack-OR. This is because GraphTrack-OR-UnNorm is significantly influenced by devices' weighted degrees, e.g., devices of heavy Internet users are biased to have large similarity scores. 
Our results show that normalizing edge weights of the IP-Device/Domain-Device graph 
can increase 1) performance when there are no errors in single-device tracking and 2) robustness to uncertainty in single-device tracking.

\myparatight{Alternatives of converting BAT-SU to its unsupervised version} We propose to convert BAT-SU to be unsupervised under our framework. Here, we explore two other alternatives to convert BAT-SU to be unsupervised.  
One way is to randomly select a threshold for BAT-SU without using a training dataset to learn it. Figure~\ref{fig:threshold} shows the Accuracy of this unsupervised version of BAT-SU as we change the threshold. 
Our results indicate that BAT-SU is sensitive to the threshold.
For instance, when the threshold is around 0.13, the Accuracy is the highest; when the threshold is 0.19, the Accuracy decreases by 21\%. 

For each mobile device, BAT-SU computes a similarity score with each desktop device. Therefore, 
another way could be to compute the gap between the largest similarity score and the second largest similarity score. If the gap is large enough (e.g., larger than a certain threshold), then BAT-SU predicts a match. 
Figure~\ref{fig:gap} shows such gaps for 20 sampled  mobile devices. For each mobile device, the desktop device with the largest similarity score correctly matches the mobile device. 
We observe that the gaps span a wide range of values, which means that it is challenging to select a gap threshold to achieve a good precision-recall tradeoff.

\begin{figure}[t]
\vspace{-4mm}
\centering
\subfloat[]{\includegraphics[width=0.24\textwidth]{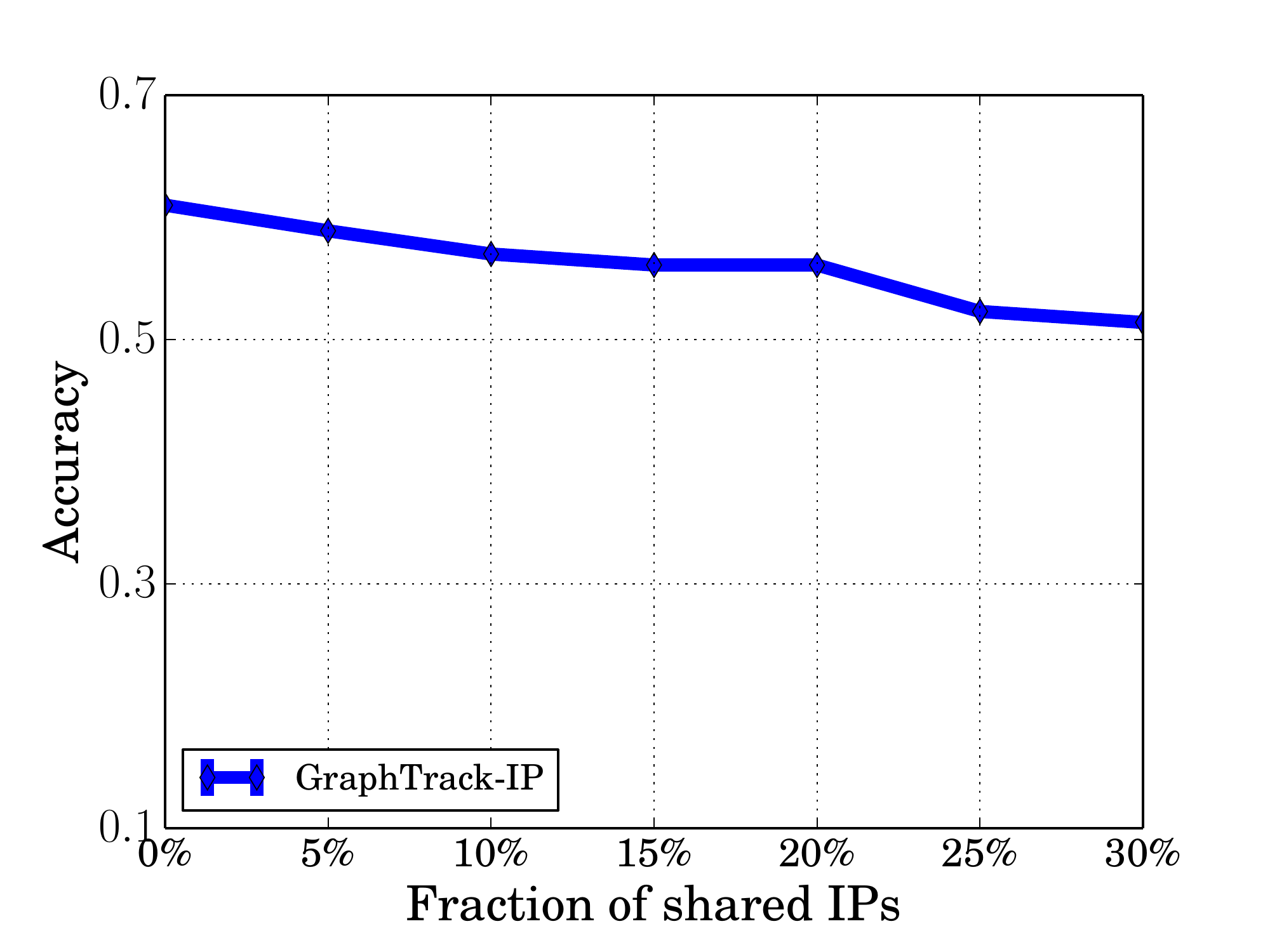} \label{result-NAPT}}
\subfloat[]{\includegraphics[width=0.24\textwidth]{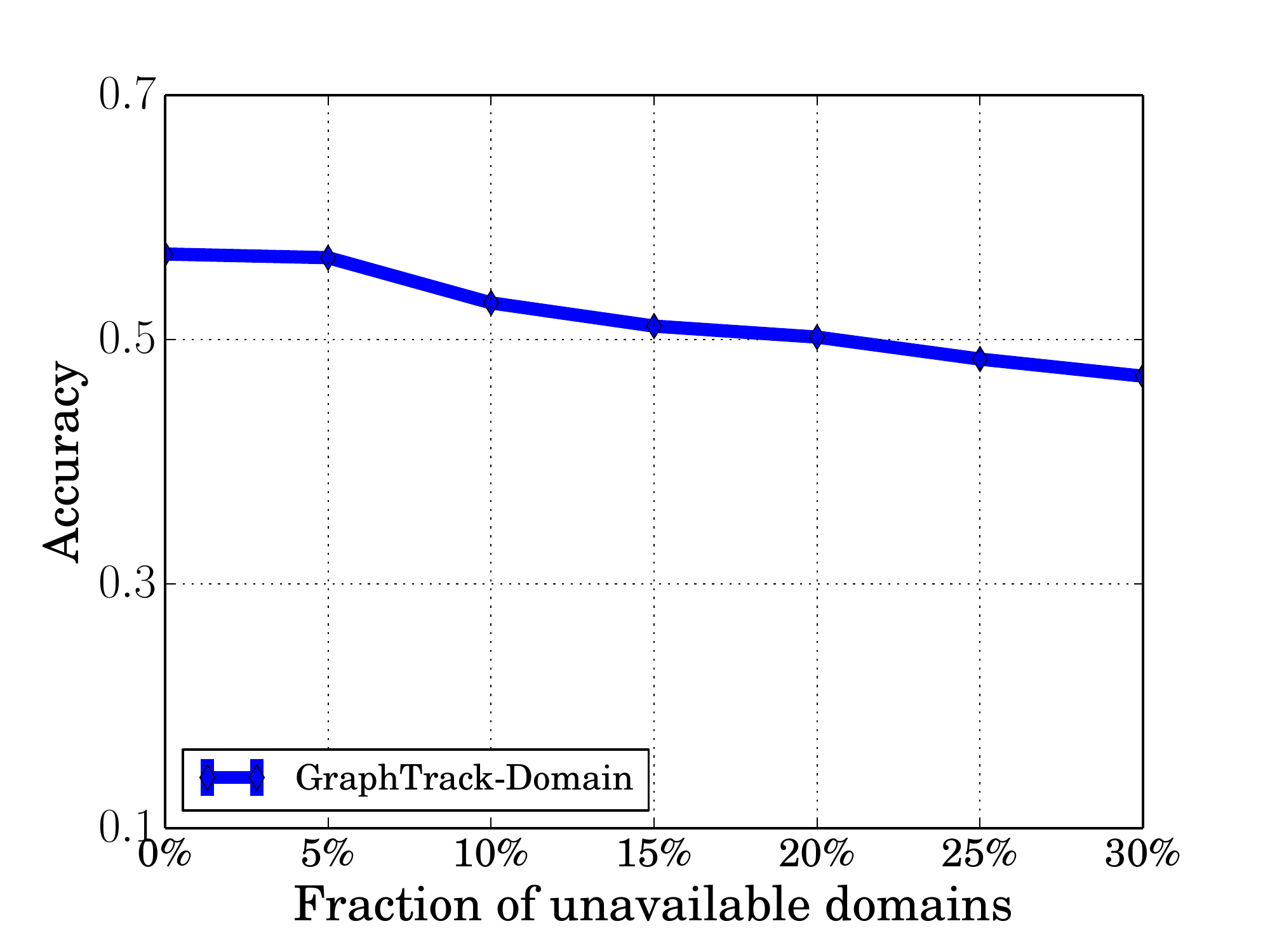} \label{result-Referer}}
\vspace{-2mm}
\caption{(a) Accuracy of GraphTrack-IP vs. fraction of shared IPs. (b) Accuracy of GraphTrack-Domain vs. fraction of unavailable domains.} 
\vspace{-6mm}
\end{figure}

\noindent {\bf Impact of shared IPs:}
In practice, the same IP may be used by different users/devices due to Network Address Port Translation (NAPT).
A natural question is how shared IPs impact the performance of GraphTrack.
To simulate shared IPs, we randomly select $x\%$ of IPs and assign each IP to some randomly selected devices (e.g., 5\% of total devices in our experiment). This means that we add edges between the selected IPs and selected devices in the IP-Device graph. Then, we calculate GraphTrack-IP's performance on the IP-Device graph  with added edges. 
Figure~\ref{result-NAPT} shows Accuracy of GraphTrack-IP vs. $x\%$ of shared IPs. We observe that GraphTrack-IP's Accuracy marginally decreases even when 30\% of IPs are shared to different devices.  

\noindent {\bf Impact of unavailable web referer:} 
In practice, a third-party tracker may be unable to know the domain a user visits, when the web referer field of browser request for a domain is not available.
To simulate this scenario, we randomly select $x\%$ of domains and assume these domains are not visited by any device. This means we remove the edges between the selected $x\%$ domains and the connected devices in the Domain-Device graph. Then, we calculate GraphTrack-Domain's performance on the Domain-Device graph with removed edges. 
Figure~\ref{result-Referer} shows Accuracy of GraphTrack-Domain vs. $x\%$ of domains are unavailable. We observe that GraphTrack-Domain's Accuracy 
only slightly decreases around 0.1 when 30\% of domains are removed in our dataset.

\subsection{Mobile-Desktop Tracking Results for Supervised Methods}

In this part, we evaluate our supervised version of GraphTrack, which would be 
used in the scenarios of a third-party tracker who also serves as the first party or a first-party tracker.   In the former case like a \texttt{Facebook like button}, the tracker could have access to labeled device pairs from multiple domains that embed the tracker. To simulate such scenarios, we randomly sample some device pairs as labeled device pairs. In the latter case like a bank website, a first-party tracker will only have access to labeled device pairs from a single domain via cross-device IDs, i.e., the tracker's domain. To simulate such scenarios, we treat device pairs that both visited a certain domain as labeled device pairs.

\begin{figure}[]
\vspace{-4mm}
\centering
\subfloat[]{\includegraphics[width=0.24\textwidth]{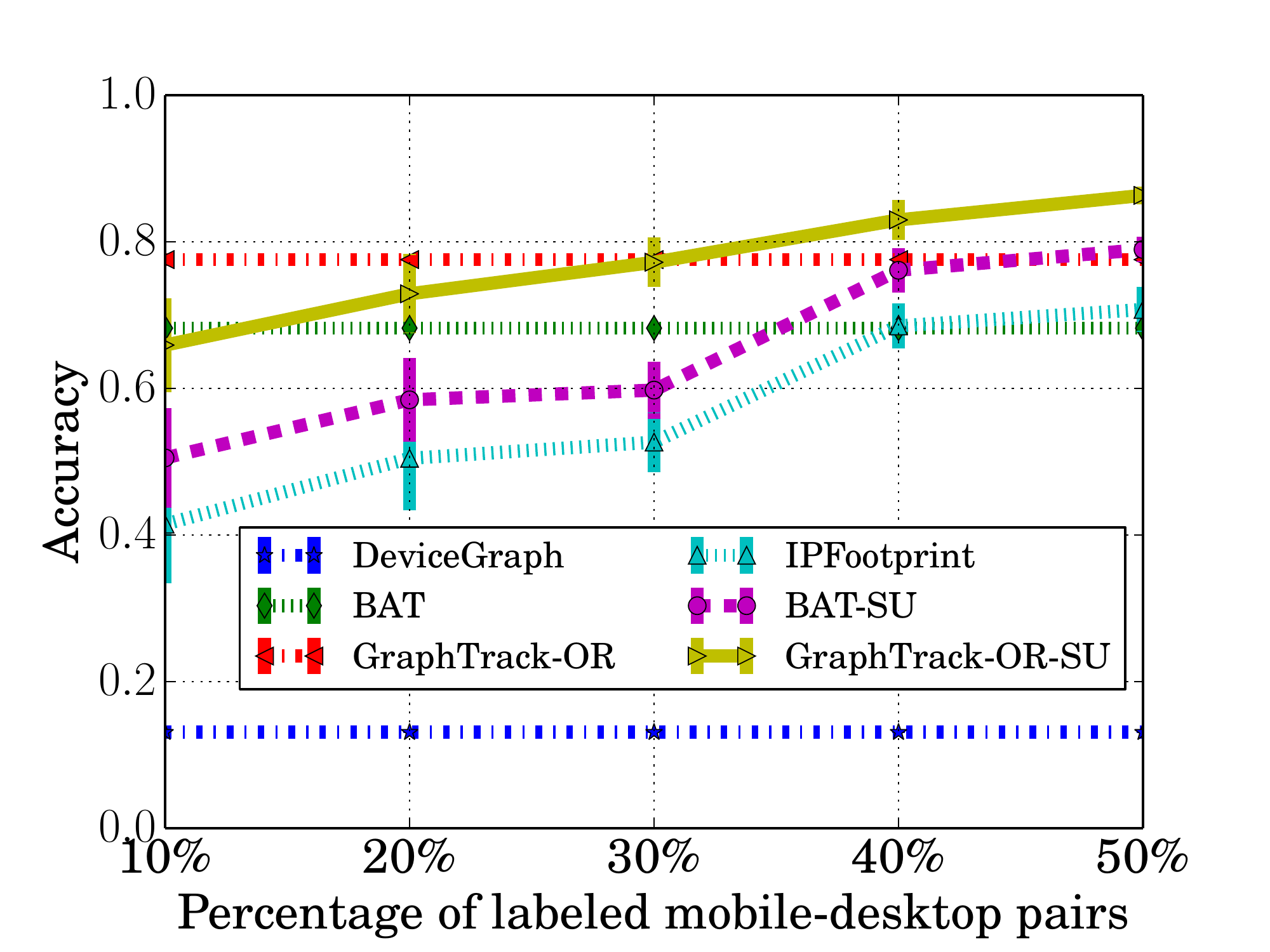} \label{result-sup-unsup-all}}
\subfloat[]{\includegraphics[width=0.24\textwidth]{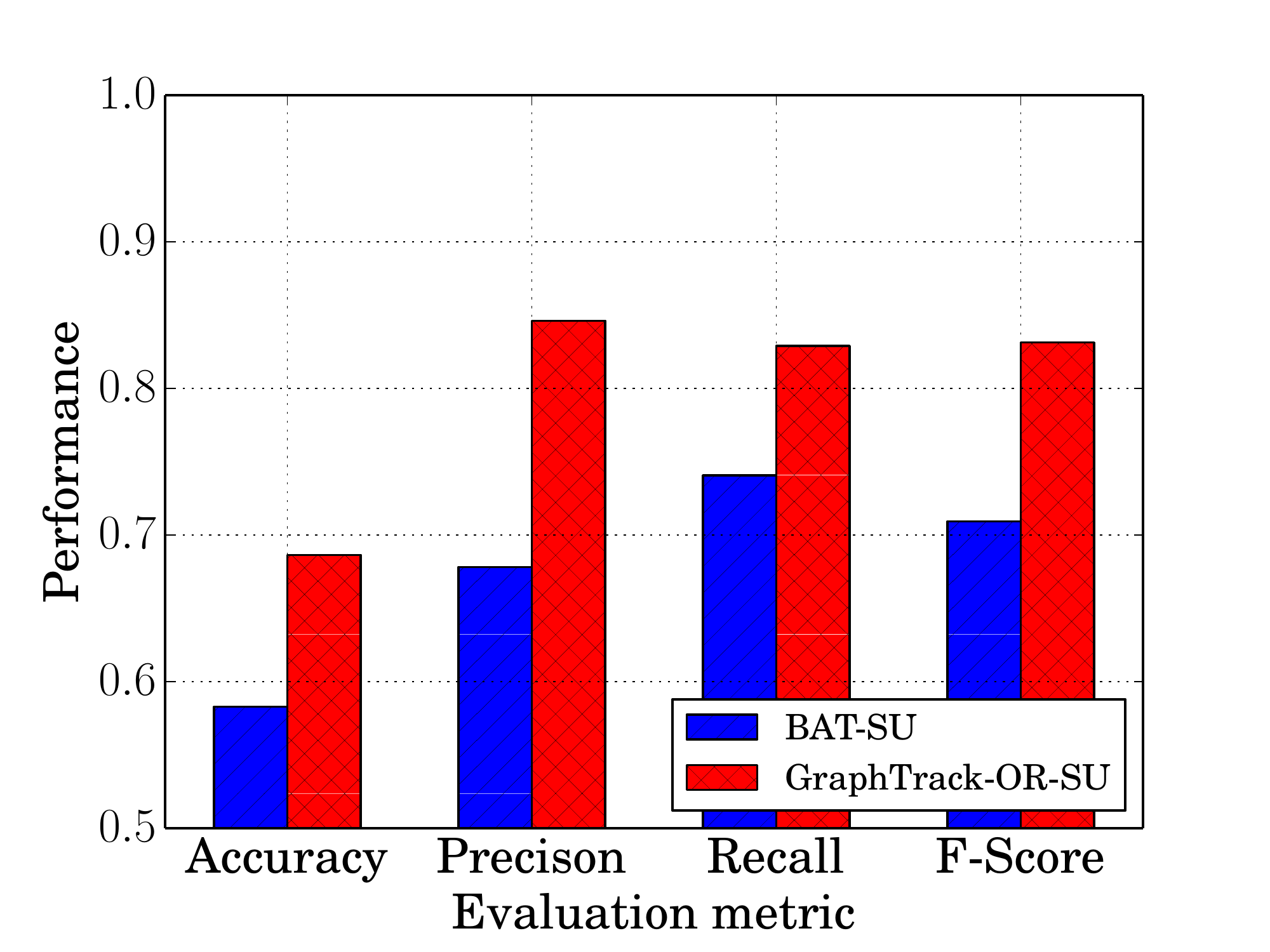} \label{result-single-domain}}
\vspace{-2mm}
\caption{Comparing supervised cross-device tracking methods, 
where labeled device pairs are (a) randomly sampled and (b) obtained via cross-device IDs on a single domain.} 
\vspace{-4mm}
\end{figure}

\noindent {\bf GraphTrack outperforms compared supervised methods:}
From Figure~\ref{result-sup-unsup-all}, we observe that GraphTrack-OR-SU significantly outperforms BAT-SU and IPFootprint. 
For instance, when we have 20\% labeled mobile-desktop pairs, GraphTrack-OR-SU's Accuracy (0.73) is around 0.15 higher than BAT-SU's (0.58) and 0.23 higher than IPFootprint's (0.50). 
Moreover, BAT-SU requires labeling 40\%-50\% of mobile-desktop pairs to outperform GraphTrack-OR. 
Our results indicate that modeling the interplays between IPs, domains, and devices using graphs is more powerful than common neighbors based similarity metrics used by BAT-SU. Moreover, tracking methods using both IPs and domains outperform methods using only IPs.

\noindent {\bf Supervised methods outperform their unsupervised counterparts with enough training samples:} Generally, supervised methods have better performance when more training samples are available. Specifically, BAT-SU, IPFootprint, and GraphTrack-OR-SU have better performance with more training samples are available. 
Moreover, a supervised method requires a large fraction of labeled pairs to outperform its unsupervised counterpart. 
For instance, GraphTrack-OR-SU outperforms GraphTrack-OR with 30\% of labeled mobile-desktop pairs, while BAT-SU outperforms 
 BAT with 30\%-40\% of labeled mobile-desktop pairs. 
 The reason is that the matched mobile-desktop pairs have diverse patterns and supervised methods require labeling a large fraction of them as a representative training set to achieve good performance.

\noindent {\bf Labeled device pairs are obtained via cross-device IDs on a single domain:} A first-party tracker could obtain labeled device pairs via cross-device IDs. In particular, when a user logs in the tracker's web service on both its desktop and mobile devices, the tracker can treat the user's mobile-desktop device pair as a labeled device pair. To simulate such a first-party tracker, we randomly select a domain in our dataset, treat it as the tracker's domain,  take the mobile-desktop pairs that both visited the domain as labeled device pairs,  and use the remaining devices to evaluate GraphTrack-OR-SU and BAT-SU. 
{Note that IPFootprint is not applicable in this scenario.}  
Figure~\ref{result-single-domain} shows the comparison results, where we repeat the experiments five times on five randomly selected domains and report the average results. We observe that GraphTrack-OR-SU also substantially outperforms BAT-SU.

\begin{figure*}[t]
\vspace{-2mm}
\centering
\subfloat[]{\includegraphics[width=0.3\textwidth]{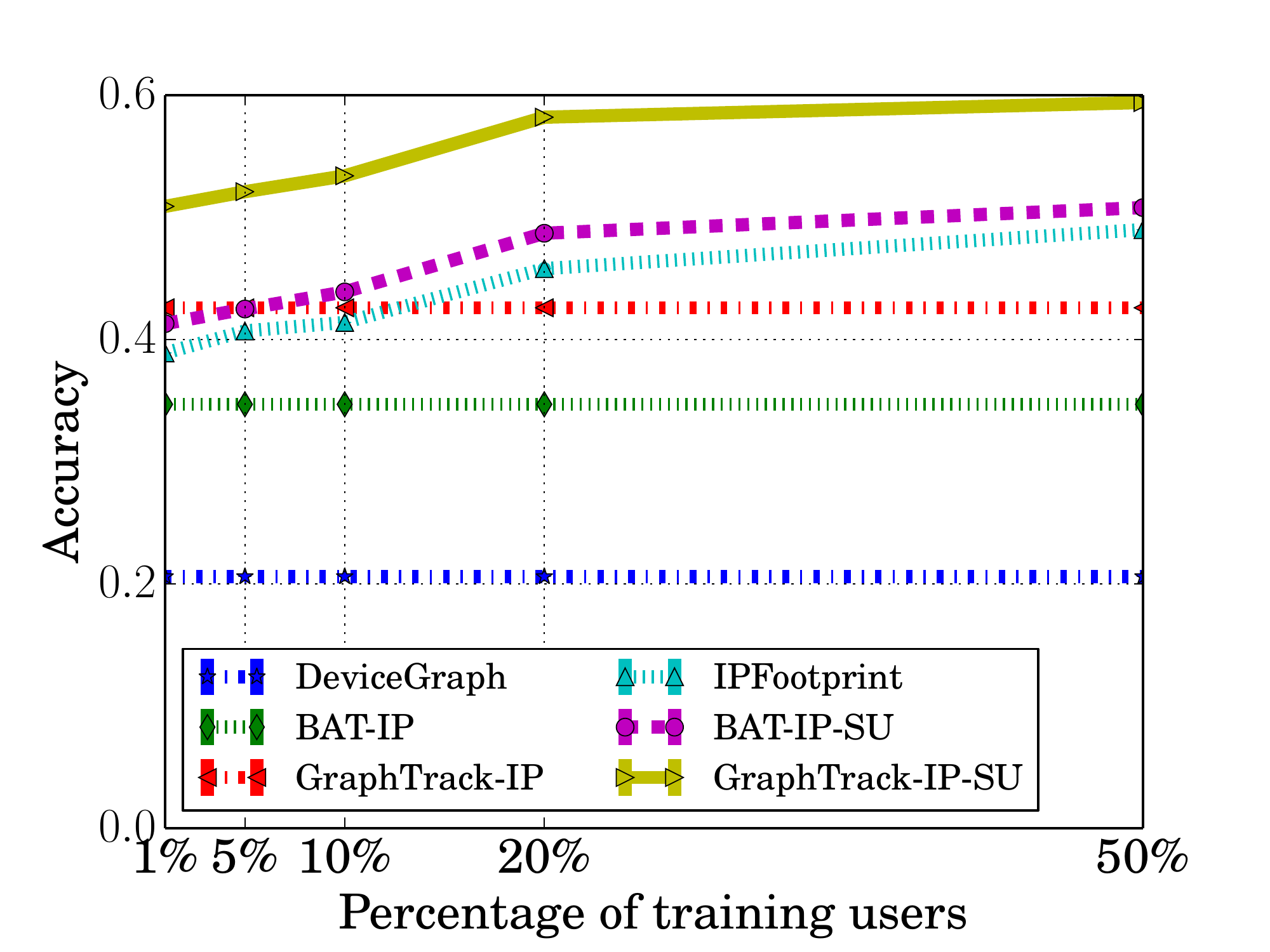} \label{result-sup-unsup-JHU} }
\subfloat[]{\includegraphics[width=0.3\textwidth]{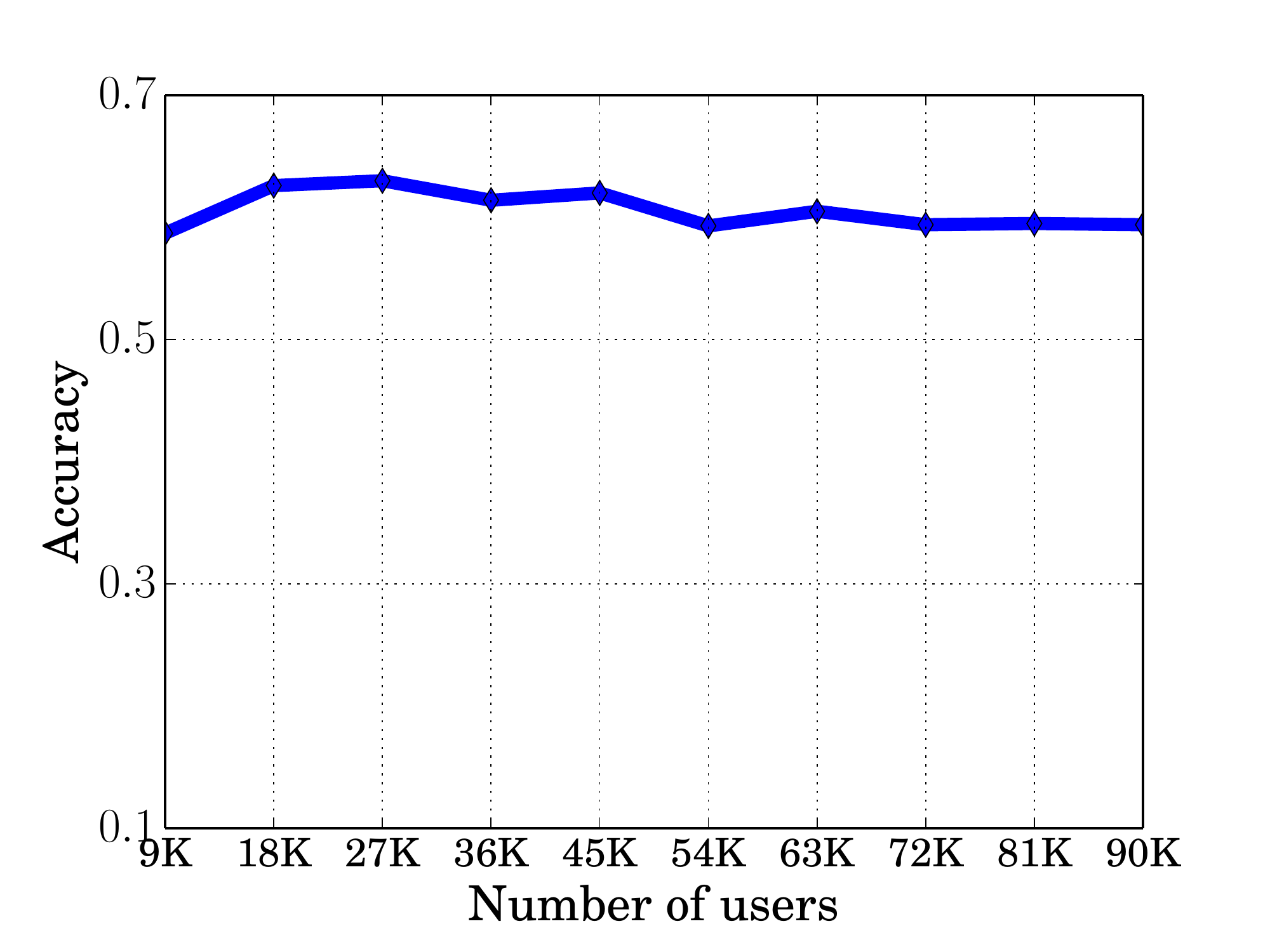} \label{impact-users} }
\subfloat[]{\includegraphics[width=0.3\textwidth]{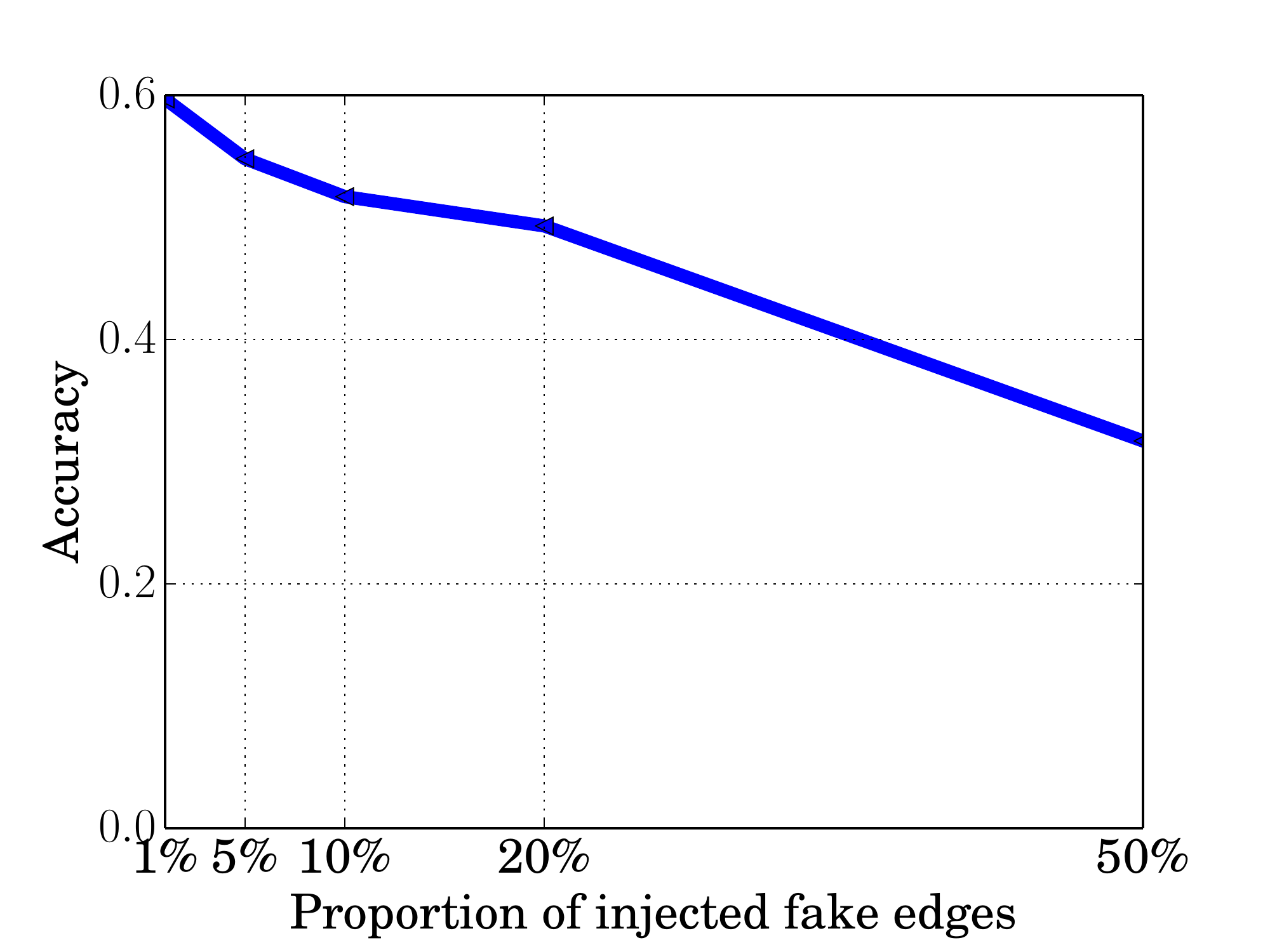} \label{defense-graphtrack-JHU}}
\vspace{-2mm}
\caption{(a) Accuracy of 
multiple-device tracking vs. percentage of randomly sampled users as training set.
(b) GraphTrack-IP-SU's accuracy for multiple-device tracking vs. number of users sampled from our multiple-device dataset. (c) GraphTrack-IP-SU's accuracy for multiple-device tracking vs. proportion of injected fake edges into the IP-Device Graph.} 
\end{figure*}

\subsection{Multiple-Device Tracking Results}
In this part, we simulate real-world cross-device tracking using our collected dataset. 
Our experiment in this scenario corresponds to a first-party tracker and only IP prefixes are used for tracking. 
We adopt our GraphTrack methods to simulate the real-world tracker. 
Specifically, we leverage the devices collected between December 2017 and March 2018 to construct the IP-Device graph; and we use 10 threads to run our methods in this graph in parallel.
Then, we dynamically update the IP-Device graph and perform our GraphTrack sequentially to link each incoming device
between April 2018 and July 2018 in our dataset.

\noindent {\bf GraphTrack outperforms compared methods:} 
We compare GraphTrack with both unsupervised and supervised methods for multiple-device tracking.
Figure~\ref{result-sup-unsup-JHU} shows the Accuracy of the compared methods. We observe that GraphTrack-IP-SU significantly outperforms IPFootprint and BAT-IP-SU. For instance, when we use 20\% users' devices for training, GraphTrack-IP-SU's Accuracy (0.59) is 0.13 and 0.1 higher than IPFootprint's (0.46) and BAT-IP-SU's (0.49), respectively. Likewise, when no labeled device groups are available, GraphTrack-IP's Accuracy (0.43) is 0.22 and 0.08 larger than DeviceGraph's (0.21) and BAT-IP's (0.35), respectively.

Table~\ref{tracking_detail} shows the Accuracy for tracking users who have a certain number of devices. 
We observe that GraphTrack does track users with more than 2 devices and consistently outperforms the compared methods for tracking multiple devices. 
Moreover, GraphTrack-IP and  GraphTrack-IP-SU outperform the compared unsupervised methods and supervised methods, respectively. 
For instance, Graph
Track-IP outperforms BAT-IP by 0.07, 0.14, 0.15, and 0.18 and outperforms DeviceGraph by 0.23, 0.22, 0.27, 0.19; 
and GraphTrack-IP-SU outperforms BAT-IP-SU by 0.09, 0.08, 0.14, and 0.05, and outperforms IPFootprint by 0.11, 0.09, 0.17, and 0.1, for tracking users with 2, 3, 4, and 5 devices, respectively.

\noindent {\bf Impact of the number of users:} 
We study the impact of the total number of users on GraphTrack's performance. In particular, we randomly sample some users between December 2017 and March 2018, 
treat 50\% of them as training users, and test GraphTrack-IP-SU on the remaining sampled users.  
Figure~\ref{impact-users} shows Accuracy of GraphTrack-IP-SU vs. number of users sampled from our multiple-device dataset.
We observe that GraphTrack-IP-SU's performance is stable with respect to the number of users. The reason is that 
GraphTrack naturally divides devices into geographically separated clusters using RWwR and each cluster has a small number of users.

\noindent {\bf Practicability of GraphTrack as a real-world tracker:}
 We study how practical GraphTrack is to link different devices using our real-world dataset collected from a real-world website from December 2017 to July 2018.  Specifically, we measure the average linking time per device as shown in Table~\ref{real_time}: GraphTrack-IP takes around 40ms and GraphTrack-IP-SU around 30ms to link an incoming device.  This is sufficient and practical for a real-world scenario, such as real-time bidding (RTB): For example, the deadline of RTB as required by Google~\cite{RTB} is between 120ms and 300ms.  Another thing worth noting here is that  BAT and BAT-SU are faster than GraphTrack-IP and GraphTrack-IP-SU in linking, which take around 15ms and 10ms respectively.  The reason is that GraphTrack is an iterative method but BAT is not; however, we would argue that GraphTrack is more accurate and the runtime latency is sufficient for the real-world website used in our experiment for the linking purpose.

\begin{table}[!t]\renewcommand{\arraystretch}{1.3}
\small
\centering
\caption{Accuracy for tracking users who have a particular number of devices, where 50\% of randomly selected users are used as training dataset for the supervised methods.}
\vspace{-3mm}
\addtolength{\tabcolsep}{-2pt}
\begin{tabular}{|c|c|c|c|c|}
\hline
 {\bf Method} & {\bf 2 devices} & {\bf 3 devices} & {\bf 4 devices} & {\bf 5 devices} \\ \hline
  {\bf DeviceGraph}  	& 0.21 & 0.13 & 0.19 & 0.06 \\ \hline
  {\bf BAT-IP}  		& 0.37 & 0.21 & 0.31 & 0.07 \\ \hline
  {\bf GraphTrack-IP} 	& {\bf 0.44} & {\bf 0.35} & {\bf 0.46} & {\bf 0.25}	 \\ \hline \hline
  {\bf IPFootprint}  	& 0.50 & 0.36 & 0.48 & 0.31 \\ \hline
  {\bf BAT-IP-SU}  		& 0.52 & 0.39 & 0.51 & 0.36 \\ \hline
 {\bf GraphTrack-IP-SU} & {\bf 0.61} & {\bf 0.47} & {\bf 0.65} & {\bf 0.41} \\ \hline
\end{tabular}
\label{tracking_detail}
\vspace{-4mm}
\end{table}

\noindent {\bf Robustness of GraphTrack to single-device tracking errors, e.g., device cookie clearance:} 
 We show how a less accurate ground-truth may affect the performance of GraphTrack.  Specifically, we use cookies---which are often cleared by users manually or even browsers automatically (like intelligent tracking prevention)---in our multi-device tracking dataset as the ground truth and evaluate the performance of GraphTrack and BAT. 
 Here are the results.  
GraphTrack-IP-SU's accuracy decreases from 0.60 with device ID as the ground truth to 0.56 with cookies; as a comparison, 
BAT-SU's accuracy decreases from 0.50 to 0.42.

\noindent {\bf Defense against GraphTrack:} 
GraphTrack has shown robustness to uncertainty in single-device tracking which incorrectly links some IPs and domains to  \emph{randomly} selected wrong devices.  
To defend against cross-device tracking, one possible strategy is that a user visits domains from some random IPs (e.g., using Tor or VPN) or visits some random domains, to decrease the similarity between his devices. To simulate such a defense, we consider the multiple-device tracking dataset and randomly inject fake edges between devices and IPs into the IP-Device graph. 
We denote $y\%$ as the ratio between the number of injected fake edges and the number of true edges in the original IP-Device graph. 
 Figure~\ref{defense-graphtrack-JHU} shows the Accuracy of GraphTrack-IP-SU, 
where 50\% of users are used as the training dataset. 
We observe that when injecting more fake edges (i.e., visiting domains using more random IPs), 
GraphTrack-IP-SU has a lower Accuracy. 
However, to significantly reduce 
the Accuracy, a user needs to visit a large number 
(e.g., 50\%) of random IPs on average.

\begin{table}[!tbp]\renewcommand{\arraystretch}{1.3}
\small
\centering
\caption{Total time and time per device of our methods to track users' devices 
in our multiple-device tracking dataset.
}
\vspace{-3mm}
\addtolength{\tabcolsep}{-4pt}
\begin{tabular}{|c|c|c|c|c|c|c|}
\hline
\multicolumn{2}{|c|}{\bf Date: 2017.12--}  & {\bf 2018.3}	& {\bf 2018.4} & {\bf 2018.5} & {\bf 2018.6} & {\bf 2018.7} \\ \hline
 \multicolumn{2}{|c|}{\bf \#Devices}&  {\bf 165K} & {\bf 208K} & {\bf 251K} & {\bf 293K} & {\bf 335K} \\ \hline
 \multirow{2}{*}{\bf GraphTrack-IP} &{\bf Total time}	& 1.8h &	2.2h & 2.6h	& 3.0h	& 3.4h \\ \cline{2-7}
 & {\bf Time per dev.} & 39ms & 38ms & 37ms & 37ms & 37ms \\ \hline
  \multirow{2}{*}{\bf GraphTrack-IP-SU} & {\bf Total time} & 1.0h & 1.4h & 1.8h & 2.2h & 2.6h \\ \cline{2-7}
  & {\bf Time per dev.} & 22ms & 24ms & 26ms & 27ms & 28ms \\ \hline
 \end{tabular}
\label{real_time}
\vspace{-4mm}
\end{table}

\section{Conclusion and Future Work}

We propose \emph{GraphTrack}, a graph-based framework, to perform cross-device tracking using browsing history. 
Specifically, we model the complex interplays between IPs, domains, and devices as an IP-Device graph and a Domain-Device graph.  
Furthermore, we adapt standard RWwR to analyze the structure of the graphs, through which we model similarity between devices and 
capture latent correlations among IPs and domains.
GraphTrack is  unsupervised, but can incorporate manual labels if available. 
GraphTrack can also be easily parallelized.
We compare GraphTrack with state-of-the-art unsupervised methods and supervised methods on two real-world datasets and demonstrate that GraphTrack substantially outperforms the compared methods.  
We conclude that graph is a more powerful tool to model relationships between heterogeneous data sources. 
Interesting future works include defending against GraphTrack via adversarial machine learning, 
studying the distribution of real-world errors for single-device trackers, and normalizing edge weights based on duration/interval.  

\noindent {\bf Acknowledgements.} We thank the anonymous reviewers for their constructive comments. This work is supported by the National
Science Foundation under Grants No. 1937787, CNS18-54000, and  and CNS18-54001. Any
opinions, findings and conclusions or recommendations expressed
in this material are those of the author(s) and do not necessarily
reflect the views of the funding agencies.

{
\bibliographystyle{ACM-Reference-Format}
\bibliography{refs,cao,websec}
}

\appendix


\begin{figure}[t]
\centering
{\includegraphics[width=0.35\textwidth]{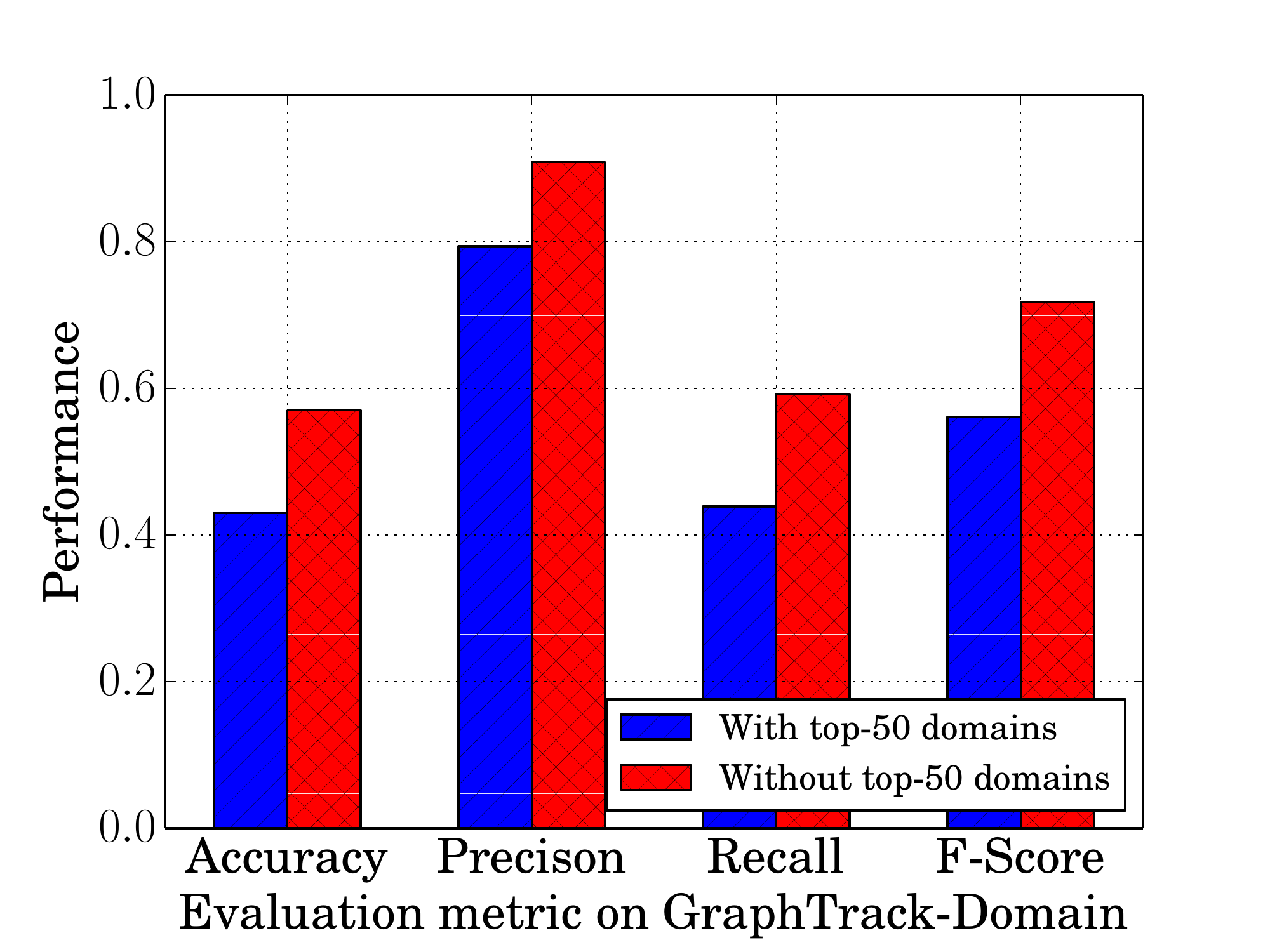}}
\caption{Performance of GraphTrack-Domain with and without the top-50 domains.}
\label{result-Domain-All}
\end{figure}

\section{Compared methods}
\label{app:comp}
We compare GraphTrack with supervised and unsupervised cross-device tracking methods. 
Table~\ref{eval_set} in Appendix summarizes all compared methods. 

{\bf Supervised methods.} We compare our supervised GraphTrack with two supervised methods, i.e., BAT-SU~\cite{zimmeck2017privacy} and IPFootprint~\cite{cao2015recovering}. 
BAT-SU~\cite{zimmeck2017privacy} is the state-of-the-art supervised method. It uses \emph{Bhattacharyya coefficient}~\cite{wang2013uniquely} to compute the similarity score between two devices. 
 BAT-SU essentially computes the weighted common neighbors between two devices in the IP-Device graph or the Domain-Device graph, where the weight is the normalized frequency of the corresponding IP or domain. 
We implement BAT-SU by ourselves and verify that our implementation achieves very close performance with the results reported by Zimmeck et al.~\cite{zimmeck2017privacy} under the same setting (See Table~\ref{result-all-metrics}).
 Note that BAT-SU is originally designed to handle 2 devices, i.e., predict a matched desktop device for a mobile device, while our multiple-device dataset includes users having more than 2 devices. Thus, we adapt BAT-SU to handle multiple devices using our GraphTrack-SU framework (Section~\ref{graphtrack_sup}). The idea is to replace the similarity scores computed by RWwR as the Bhattacharyya coefficient.     

IPFootprint~\cite{cao2015recovering} only uses  IPs to compute the similarity score between two devices. 
It represents each device's IPs as a feature vector, where the $i$-th entry is the frequency the device used the $i$-th IP. 
Each IP is also associated with an importance parameter to be learnt. 
Then, the similarity score between two devices is defined as the dot product of two devices' feature vectors and importance vectors. 
The importance vector is learnt by a learning to rank method called RankNet~\cite{burges2005learning}. Specifically, given a training set consisting of groundtruth matched 
and unmatched device pairs, the learnt ranking model outputs the likelihood that each device matches any other device.
Then, IPFootprint designs an ensemble method to determine the devices predicted to match a given device.

{\bf Unsupervised methods.} We compare our unsupervised GraphTrack with BAT and DeviceGraph~\cite{deviceGraphKDD17}. BAT is an adapted unsupervised version of BAT-SU using our GraphTrack framework.
Specifically, BAT replaces the similarity scores in GraphTrack-OR with the Bhattacharyya coefficients. We use the OR operator because, as we will demonstrate, it outperforms the AND operator and the unified graph at combining IPs and domains. 

DeviceGraph~\cite{deviceGraphKDD17} constructs a \emph{device graph} based on IP co-locations and leverages community detection to track users' devices. Specifically, a node in the device graph is a device and an edge is created between two devices if they used the same IP. Then, DeviceGraph leverages Louvain method~\cite{blondel2008fast} to detect communities in the device graph and predicts that devices in a community belong to the same user. We note that DeviceGraph only leverages IPs.

\begin{table}[!t]\renewcommand{\arraystretch}{1.5}
\small
\centering
\caption{Results of BAT-Raw, our implemented BAT-SU, IPFootprint, and our GraphTrack-OR-SU on the same training set and testing set as in~\cite{zimmeck2017privacy}. 
BAT-Raw has 37 TPs, 5 FPs, 0 TN, and 2 FNs; BAT-SU has 36 TPs, 6 FPs, 0 TN, and 2 FNs; IPFootprint has 29 TPs, 6 FPs, 0 TN, and 9 FNs; GraphTrack-OR-SU has 39 TPs, 2 FPs, 0 TN, and 3 FNs.
}
\addtolength{\tabcolsep}{-3pt}
\begin{tabular}{|c|c|c|c|c|}
\hline
 &  {\bf Accuracy} & {\bf Precision} & {\bf Recall} & {\bf F-Score} \\ \hline
 {\bf BAT-Raw} & 0.84 & 0.88 & 0.95 & 0.91 \\ \hline
 {\bf BAR-SU} & 0.82 & 0.86 & 0.95 & 0.90 \\ \hline
 {\bf IPFootprint} & 0.66 & 0.83 & 0.76 & 0.79\\ \hline 
 {\bf GraphTrack-OR-SU} & 0.89 & 0.95 & 0.93 & 0.94\\ \hline
\end{tabular}
\label{result-all-metrics}
\end{table}

\section{Verifying our Implementation}
\label{reproduce}

We denote the implementation of BAT-SU by its authors~\cite{zimmeck2017privacy} as BAT-Raw. 
We verify that our implementation of BAT-SU reproduces BAT-Raw. 
Specifically, we obtained the same training set and testing set from the authors~\cite{zimmeck2017privacy}. 
The training set consists of 63 matched mobile-desktop pairs and the remaining 44 matched pairs form the testing set. Table~\ref{result-all-metrics} shows Accuracy, Precision, Recall, and F-Score of BAT-Raw, BAT-SU, and GraphTrack-OR-SU.
According to~\cite{zimmeck2017privacy}, BAT-Raw has 37 TPs, 5 FPs, 0 TN, and 2 FNs. 
Our implemented BAT-SU has 36 TPs, 6 FPs, 0 TN, and 2 FNs.
Our results are not exactly the same as BAT-Raw because BAT-Raw filtered the top-50 domains ranked by Alexa and all columbia.edu domains which are anonymized by cryptographic hashing in the released dataset, while we filtered the top-50 most popular domains in the dataset instead.
For comparison, our GraphTrack-OR-SU has 39 TPs, 2 FPs, 0 TN, and 3 FNs. 
Our GraphTrack-OR-SU outperforms BAT-SU for supervised cross-device tracking.

\begin{table*}[!t]\renewcommand{\arraystretch}{1.3}
\footnotesize
\caption{Evaluation settings of methods used in our paper. }
\addtolength{\tabcolsep}{-5pt}
\begin{tabular}{|c|c|c|c|c|c|}
\hline
\multicolumn{2}{|c|}{\bf Methods} & {\bf IP} & {\bf Domain} & {\bf IP+Domain} & {\bf Description} \\ \hline
\multirow{11}{*}{\bf {Unsupervised methods}}  & {\bf GraphTrack-IP}  & \checkmark &  &  & RWwR on normalized weighted IP-Device graph \\ \cline{2-6}

					& {\bf GraphTrack-Domain}  &  & \checkmark &  & RWwR on normalized weighted Domain-Device graph \\ \cline{2-6} 
					& {\bf GraphTrack-UniGraph}  &  &  & \checkmark & RWwR on normalized weighted IP-Domain-Device graph \\ \cline{2-6} 
					& {\bf GraphTrack-AND}  &  &  & \checkmark &  GraphTrack-IP \emph{AND} GraphTrack-Domain  \\ \cline{2-6} 
					& {\bf GraphTrack-OR}  &  &  & \checkmark &   GraphTrack-IP \emph{OR} GraphTrack-Domain \\ \cline{2-6} 
                   & {\bf GraphTrack-OR-UnWeighted}  &  &  & \checkmark &  GraphTrack-OR with unweighted IP-Device graph/Domain-Device graph \\ \cline{2-6} 
                   & {\bf GraphTrack-OR-UnNorm}  &  &  & \checkmark &  GraphTrack-OR with unnormalized weighted IP-Device graph/Domain-Device graph \\ \cline{2-6} 
                   & {\bf BAT-IP}  & \checkmark &  &  &  Bhattacharyya coefficient on weighted IP-Device graph \\ \cline{2-6}
                   & {\bf BAT-Domain}  & \checkmark &  &  &  Bhattacharyya coefficient on weighted Domain-Device graph \\ \cline{2-6}
                   & {\bf BAT}  &  &  & \checkmark &  BAT-IP \emph{OR} BAT-Domain  \\ \cline{2-6} 
                   & {\bf DeviceGraph}  & \checkmark &  &  & Community detection on weighted device graph construted using IPs\\ \hline \hline
\multirow{9}{*}{\bf {Supervised methods}}   & {\bf GraphTrack-IP-SU}  & \checkmark &  &  &  Supervised version of GraphTrack-IP \\ \cline{2-6} 
                   & {\bf GraphTrack-Domain-SU}  &  & \checkmark &  & Supervised version of GraphTrack-Domain \\ \cline{2-6} 
                    & {\bf GraphTrack-UniGraph-SU}  &  &  & \checkmark &  Supervised version of GraphTrack-UniGraph \\ \cline{2-6} 
                    & {\bf GraphTrack-AND-SU}  &  &  & \checkmark & Supervised version of GraphTrack-AND \\ \cline{2-6} 
                    & {\bf GraphTrack-OR-SU}  &  &  & \checkmark &  Supervised version of GraphTrack-OR \\ \cline{2-6} 
                    & {\bf BAT-IP-SU}  &  \checkmark &  & &  Supervised version of BAT-IP \\ \cline{2-6}
                    & {\bf BAT-Domain-SU}  &   & \checkmark & & Supervised version of BAT-Domain \\ \cline{2-6}
                   & {\bf BAT-SU}  &  &  & \checkmark &  Supervised version of BAT \\ \cline{2-6}
                   & {\bf IPFootprint}  & \checkmark &  &  &  RankNet on pairwise devices' IP footprints \\ \hline
\end{tabular}
\label{eval_set}
\end{table*}

\section{Computational Complexity}
\label{comp}

\myparatight{Complexity of unsupervised GraphTrack methods}
We denote the number of nodes of the IP-Device graph, Domain-Device graph, and IP-Device-Domain graph as $|V|_{IP}$, $|V|_{DO}$, and $|V|_{Uni}$, respectively. Moreover, we denote the number of edges of the three graphs as $|E|_{IP}$, $|E|_{DO}$, and $|E|_{Uni}$, respectively. 
Note that $|E|_{Uni} = |E|_{IP} + |E|_{DO}$.
For simplicity, we take GraphTrack-IP as an example, since other unsupervised methods share the same analysis.
 {Suppose we have $n$ devices $\mathcal{D}= \{D_1, D_2, \cdots, D_n \} $.}

In {Step I}, GraphTrack-IP starts an RWwR from each device $D_i$ in the IP-Device graph 
and computes the similarity scores between $D_i$ and other devices using the stationary distribution of the RWwR. 
Usually, it is sufficient to run RWwR $\log |V|_{IP}$ iterations to reach the stationary distribution~\cite{tong2006fast}, and each iteration traverses all edges of the IP-Device graph. Thus, the time complexity of performing an RWwR for $D_i$ is $O(\log |V|_{IP} \cdot |E|_{IP})$.
Next, GraphTrack-IP finds the $K-1$ candidate devices having the largest similarity scores with $D_i$. This is implemented by first sorting the similarity scores associated with the $n$ devices 
in a descending order and then selecting the top-$(K-1)$ indexes in the sorted similarity scores as the $K-1$ candidate devices. Sorting the similarity scores of $n$ devices has a time complexity $O(n \cdot \log n)$
and the time complexity for top-$(K-1)$ devices selection can be ignored. Thus, the time complexity of finding the $K-1$ candidate devices for $D_i$ is $O(n \cdot \log n)$. 
Repeating above RWwR and candidate devices selection for $n$ devices, 
we have the time complexity $O\big(n \cdot \log |V|_{IP} \cdot|E|_{IP} + n^2 \cdot \log n \big)$. 
{Step II} is device similarity graph construction using the candidate devices of each device and {Step III} is a simple prediction for each device. Both their time complexity can be ignored. 
Therefore, we have the overall time complexity of GraphTrack-IP as $O\big(n \cdot \log |V|_{IP} \cdot|E|_{IP} + n^2 \cdot \log n \big)$.

Likewise, time complexity of GraphTrack-Domain and GraphTrack-UniGraph are $O\big(n \cdot \log |V|_{DO} \cdot|E|_{DO} +  n^2 \cdot \log n \big)$ and $O\big(n \cdot \log |V|_{Uni} \cdot|E|_{Uni} + n^2 \cdot \log n \big)$; 
and GraphTrack-OR and GraphTrack-AND have the same time complexity $O\big(n \cdot (\log |V|_{DO} \cdot|E|_{DO} + \log |V|_{IP} \cdot|E|_{IP}) + 2n^2 \cdot \log n \big)$.

\myparatight{Complexity of supervised GraphTrack methods}
 {
We denote the number of labeled devices for training as $n_{tr}$, and the number of devices for testing as $n - n_{tr}$. 
Similarly, we take GraphTrack-IP-SU as an example, since other supervised methods share the same analysis.
 
GraphTrack-IP-SU first uses $n_{tr}$ labeled devices in the training set to learn a threshold. 
Specifically, for each device $D_j$ in the training set, GraphTrack-IP-SU first starts an RWwR from $D_j$ and computes the stationary distribution of the RWwR among the IP-Device graph in $\log |V|_{IP}$ iterations. The time complexity is $O(\log |V|_{IP} \cdot|E|_{IP})$. 
Then, GraphTrack-IP-SU uses the stationary distribution as the similarity scores between $D_j$ and other devices and finds the $(K-1)$ candidate devices for $D_j$ based on the similarity scores. 
The time complexity is $O(n \cdot \log n)$. 
Repeating above step for $n_{tr}$ training devices, we have the time complexity $O(n_{tr} \cdot (\log |V|_{IP} \cdot|E|_{IP} + n \cdot \log n))$.
Next, for each device pair in the training set, GraphTrack-IP-SU checks whether one device matches its paired device, i.e., whether one device is among the $(K-1)$ candidate devices of its paired device.
 Repeating the process for all labeled device pairs, GraphTrack-IP-SU can determine the threshold as the minimum similarity score that the device pair matches. 
The time complexity in this step can be ignored. 
Thus, the time complexity of training is $O(n_{tr} \cdot (\log |V|_{IP} \cdot |E|_{IP} + n \cdot \log n))$. 

Then, GraphTrack-IP-SU performs an RWwR from each device in the testing set and finds its candidate devices based on the stationary distribution of the RWwR. 
Repeating this step for all testing devices, we have the time complexity $O((n - n_{tr}) \cdot (\log |V|_{IP} \cdot |E|_{IP} + n \cdot \log n))$. 
Furthermore, GraphTrack-IP-SU constructs the device similarity graph 
to match testing devices 
based on the learnt threshold 
and the candidate devices of each testing device. The time complexity in this step can be ignored.   
Thus, the overall time complexity of GraphTrack-IP-SU is $O(n \cdot \log |V|_{IP} \cdot|E|_{IP} + n^2 \cdot \log n)$, 
the same time complexity as GraphTrack-IP.}
Likewise, GraphTrack-Domain-SU, GraphTrack-UniGraph-SU, GraphTrack-OR-SU, and GraphTrack-AND-SU also have the same time 
complexities as their unsupervised versions.

\myparatight{Speeding up GraphTrack methods} 
We have a two-level parallel implementation to speed up GraphTrack methods on large-scale datasets. 
First, different target devices can be run on different machines, as these target devices are matched in sequence. 
Second, each machine can parallelize GraphTrack using multithreading. 
Specifically, we first divide nodes in a graph into groups. Then in each iteration, each thread applies our adapted RWwR in Equation~\ref{computeP} to a group of nodes and the probability distributions obtained by all threads are combined to form the probability distribution.

\section{Discussion}
\label{discussion}

\noindent {\bf Advanced defense against GraphTrack} 
A potential better strategy to defend against GraphTrack is to leverage adversarial machine learning, which may be a valuable future work.
For instance, we can convert the problem of injecting fake edges between devices and IPs/domains into an optimization problem, where the objective function is to minimize the GraphTrack's performance
 (i.e., minimizes the correct matches between mobile devices and desktop devices) 
 and the constraint restricts the ratio of injected edges. Then, the solution to the optimization problem is
 specifically crafted fake edges 
 that minimize the GraphTrack's performance.

\noindent {\bf Different ways to normalize the edge weights:}	
We normalized the edge weight between a device and an IP/Domain based on the frequency the device accessed/visited the IP/Domain. 
Considering duration/interval that each device accessed/visited IPs/Domains 
to \vfill\eject 
\noindent  assign edge weights may further improve GraphTrack's performance. 
Our datasets do not have such information, but we believe it would be a valuable future work to consider such information.

\noindent {\bf Graph embedding methods for cross-device tracking:}
Algorithmically, graph embedding methods, e.g., Deepwalk~\cite{perozzi2014deepwalk}, node
2vec~\cite{grover2016node2vec}, struc2vec~\cite{ribeiro2017struc2vec}, can be also used to perform cross-device tracking. However, we did not choose these methods because of two reasons.
(i) \emph{Efficiency}: Graph embedding methods need to first learn feature representations for each node, which is computationally intensive.
(ii) \emph{Incremental Update}: Graph embedding methods cannot incrementally update the node feature representation if the graph includes more nodes or edges.  By contrast, GraphTrack can quickly update a new probability distribution vector given a slight change to the graph within a small number of iterations.

\end{document}